\documentclass[a4paper,10pt]{article}
\pdfoutput=1

\usepackage{jcappub}

\usepackage[style=base, singlelinecheck=off, font=small, labelfont=bf, 
justification=raggedright]{caption}
\usepackage{subfigure}

\usepackage[utf8]{inputenc}

\usepackage{float}
\usepackage{bm}
\usepackage{latexsym}
\usepackage{graphicx}
\usepackage[export]{adjustbox}
\usepackage{color}
\usepackage[dvipsnames]{xcolor}
\usepackage{verbatim}
\usepackage{array}
\usepackage{todonotes}

\def\Eq#1{eq.~\eqref{#1}}

\newcommand{\ba}{\textbf{a}}
\newcommand{\bS}{\textbf{S}}
\newcommand{\bd}{\textbf{d}}
\newcommand{\bD}{\textbf{D}}
\newcommand{\bO}{\textbf{O}}
\newcommand{\bN}{\textbf{N}}
\newcommand{\bT}{\textbf{T}}
\newcommand{\bI}{\textbf{I}}
\newcommand{\bp}{\textbf{p}}

\newcommand{\calf}{\mathcal{F}}
\newcommand{\calH}{\mathcal{H}}
\newcommand{\calL}{\mathcal{L}}

\newcommand{\calP}{\mathcal{P}}
\newcommand{\pd}{\partial}

\newcommand{\htn}{\hat{n}}
\newcommand{\htp}{\hat{p}}

\title{Joint Bayesian analysis of large angular scale CMB temperature anomalies}

\author[1]{Shabbir Shaikh,} 
\emailAdd{shabbir@iucaa.in} 
\affiliation [1]{Inter University Centre for Astronomy and Astrophysics, Post Bag 4, Ganeshkhind, 
Pune-411007, India}

\author[2,3,4]{Suvodip Mukherjee,}
\emailAdd{mukherje@iap.fr}
\affiliation[2]{Institut d'Astrophysique de Paris\\ 98bis Boulevard Arago, 75014 Paris, France}
\affiliation[3]{Sorbonne Universités, Institut Lagrange de Paris \\ 98 bis Boulevard Arago, 75014 Paris, France}
\affiliation [4]{Centre for Computational Astrophysics, Flatiron Institute, 162 5th Avenue, 10010, 
New York, NY, USA}
\author[5,6]{Santanu Das,}
\emailAdd{sdas33@wisc.edu}
\affiliation[5]{Department of Physics, University of Wisconsin-Madison, 1150 University Avenue Madison, USA}
\affiliation[6]{Fermi National Accelerator Laboratory, PO Box 500, Batavia, IL 60510, USA}

\author[2,3,4,7]{Benjamin D. Wandelt,}
\emailAdd{bwandelt@iap.fr}
\affiliation[7]{Departments of Physics and Astronomy, University of Illinois at Urbana-Champaign, 1002 W Green St, Urbana, IL 61801, USA}

\author[1]{Tarun Souradeep}
\emailAdd{tarun@iucaa.in}

\date{\today}
\abstract{
Cosmic microwave background measurements show an agreement with the concordance cosmology model except for a few notable anomalies:  \textit{Power Suppression}, the lack of  large scale power in the temperature data compared to what is expected in the concordance model, and \textit{Cosmic Hemispherical Asymmetry}, a dipolar breakdown of statistical isotropy. An expansion of the CMB covariance in  Bipolar Spherical Harmonics naturally parametrizes both these large-scale anomalies, allowing us to perform an exhaustive, fully Bayesian joint analysis of the power spectrum and violations of statistical isotropy up to the dipole level. Our analysis sheds light on the scale dependence of the Cosmic Hemispherical Asymmetry. Assuming a scale-dependent dipole modulation model with a two-parameter power law form, we explore the posterior pdf of  amplitude $A(l = 16)$ and the power law index $\alpha$ and find the maximum \textit{a posteriori}  values  $A_*(l = 16) = 0.064 \pm 0.022$ and  $\alpha_* = -0.92 \pm 0.22$.
The maximum \textit{a posteriori} direction associated with the Cosmic Hemispherical Asymmetry is $(l,b) = (247.8^o, -19.6^o)$ in Galactic coordinates, consistent with previous analyses.
We evaluate the Bayes factor $B_{SI-DM}$ to compare the Cosmic Hemispherical Asymmetry model with the isotropic model. The data prefer but do not substantially favor the anisotropic model ($B_{SI-DM}=0.4$). We consider several priors and find that this evidence ratio is robust to prior choice. The large-scale power suppression does not soften when jointly inferring both the isotropic power spectrum and the parameters of the asymmetric model, indicating no evidence that these anomalies are coupled.}

\begin{document}

\maketitle

\section{Introduction}\label{Introduction}
The statistical homogeneity and isotropy of the universe are fundamental assumptions underlying cosmology. These assumptions have been put to the test using various cosmological probes \cite{Eriksen:2003db, Hansen:2004vq, Hoftuft-Eriksen, Planck_2015_isotropy, Planck_2013_isotropy, Hirata:2009ar, Alonso:2014xca}. Precise and almost full sky measurements of CMB temperature anisotropy provide an unique opportunity to carry out tests of statistical isotropy. As a consequence,  soon after WMAP published its first results, several studies that submitted the data to tests of statistical isotropy, uncovered hints of some anomalous features in the CMB data \cite{Eriksen:2003db, Hansen:2004vq, Hoftuft-Eriksen}. Prominent among the CMB anomalies is the Power Asymmetry or the Cosmic Hemispherical  Asymmetry (CHA), which is characterized by a $\approx 14\%$ excess of power in one hemisphere of CMB temperature fluctuations compared to the other and points to a possible challenge to the assumption of statistical isotropy in cosmology \cite{Eriksen:2003db, Hansen:2004vq}. CHA was later confirmed to also exist in the \textit{Planck} CMB map at similar amplitude as seen in the WMAP data \cite{Planck_2013_isotropy, Planck_2015_isotropy}. Further, detailed analysis enabled in the Bipolar Spherical Harmonic representation by the \textit{Planck}  collaboration confirmed its dipolar nature, frequency independence and the fact that it existed at low multipoles (large angular scales) and died off at high multipoles (small angular scales) \cite{Planck_2013_isotropy,Planck_2015_isotropy}. Various physical models have been put forth 
for the plausible origin of CHA in the CMB, see for example \cite{Erickcek:2008sm, Donoghue:2007ze, Erickcek:2009at, Ackerman:2007nb, Mukherjee:2014lea, Dai:2013kfa}. 

Apart from the presence of CHA at large-angular scales, temperature data from WMAP and \textit{Planck} also indicates lower temperature fluctuation on  large angular scales than  predicted by the concordance Lambda-Cold Dark Matter ($\Lambda$CDM) model with power law form of the power spectrum of initial density fluctuations \cite{Harrison:1969fb, Zeldovich:1972zz, 1970ApJ...162..815P, Ade:2015lrj}. An existing body of literature also explores connections between different anomalies. For example, Muir et al. (2018) \cite{Muir:2018hjv} investigate the covariance between various anomalies. Polastri et al. (2015) \cite{Polastri:2015rda} study the connection between directional anomaly of low multipole alignment and dipole modulation. Schwarz et al. (2016) \cite{Schwarz:2015cma} investigate different anomalies to find features that can explain more than one observed anomalous signal.
\begin{figure}
\centering
\includegraphics[width=0.6\linewidth]{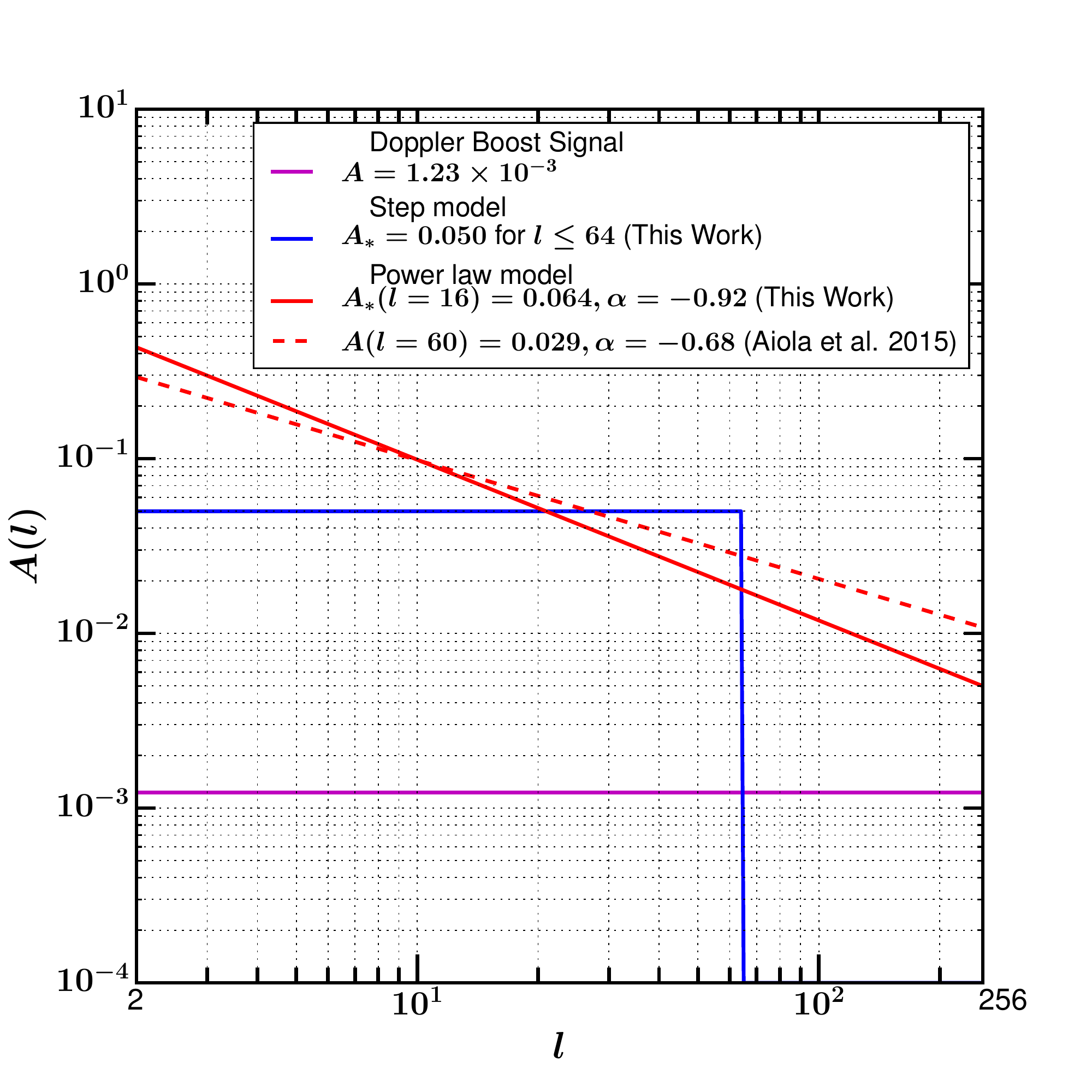}
\caption{Figure shows the dipole amplitudes as a function of multipole index $(l)$ for a phenomenological model of CHA studied here (see \Eq{eq_power_law_Al}). Magenta line depicts the amplitude of the Doppler Boost signature which is independent of the multipole 
\cite{Riess:1995cg, 2003PhRvD..67f3001K,Planck_2013_DB}. In blue and red we show the amplitude of dipole modulation for the step 
model and for the power law model respectively, obtained in this work. Solid red line shows the dipole profile which corresponds to $A_*(l_p) = 0.064$ and $\alpha = -0.92$. $A_*(l_p)$ is the value of dipole amplitude at which dipole vector has the maximum probability. The detailed analysis to obtain this  recovered scale-dependent profile is given in section \ref{results}. For comparison, we also plot the power law profile from the previous study (Aiola et al. 2015 \cite{Aiola_2015_power_law}) in red dashed line.}\label{fig_smica_Al}
\end{figure}

In this paper, we revisit the Planck-2015 temperature data \cite{Planck_2015_maps} to scrutinize both these anomalies jointly within a common mathematical framework of Bipolar Spherical Harmonics (BipoSH). BipoSH \cite{Amir_Tarun_Biposh_2005,1988qtam.book.....V} forms a complete basis for the two point function on the sphere and captures the entire structure of the covariance matrix. Hence, it is a natural choice of basis for this analysis. Any similar estimator of the two-point statistics in spherical harmonic space for a random field can be represented in the BipoSH space.
The angular power spectrum is represented by the $L = 0$ BipoSH coefficients. We explore the probability distribution of $L = 1$ (dipole) BipoSH coefficients which is the next order term in the BipoSH expansion of the covariance matrix, hence avoiding a posteriori choice. We perform the Bayesian inference of $L = 1$ BipoSH coefficients jointly with the angular power spectrum ($L = 0$ BipoSH coefficients). This is the first time a joint analysis of two anomalies in observed CMB data is being done without assuming a particular shape of the angular power spectrum.
We extend the Hamiltonian Monte Carlo (HMC) sampling method \cite{Duane:1987de,2012arXiv1206.1901N} in the BipoSH framework by Das et al. \cite{SantanuDas} to incorporate essential observational non-idealities in the Planck data 
\footnote{While this work was in progress, one of the coauthors (Santanu Das) has posted two arXiv preprints \cite{Das:2018hnr, Das:2019kvm} (PDF file of one of those was withdrawn later \cite{Das:2018hnr}) which are related to the HMC analysis for statistically anisotropic signal in the  presence of partial sky and non-isotropic noise. However, the analysis done in this paper did not use the code made public in \cite{Das:2019kvm}. The method and the code used in this analysis are developed independently from the initial code of \citep{SantanuDas} for full sky and isotropic noise.}.
We test our algorithm on simulated maps generated using \texttt{CoNIGS} algorithm \cite{Mukherjee_conigs} and then apply it to the Planck-2015 temperature data.  
Our joint estimation validates the existence of the anomalies in the \textit{Planck} temperature data and recovers a non-trivial and physically interesting profile for the scale-dependent CHA signal when modeled as a power-law form. The recovered scale-dependent CHA profile is one of the key findings of this analysis and is depicted in figure \ref{fig_smica_Al}. The details of the analysis and its significance are discussed in section \ref{results}. The obtained power-law profile sheds light onto the scale-dependent nature of the signal and goes beyond the simplistic step-model with an arbitrarily chosen cut-off \cite{Hansen:2004vq, Hoftuft-Eriksen, Planck_2013_isotropy, Planck_2015_isotropy} as shown by the blue curve in figure \ref{fig_smica_Al}. For comparison, we also plot the scale-independent signal due to our local motion \cite{Planck_2013_DB} which also exhibits a dipolar anisotropy in the CMB data. We use the parameter samples drawn from the probability distribution to perform the model comparison between statistically isotropic (SI) model and power law dipole modulation model of CHA using the ratio of Bayesian evidence obtained under two models. We note that these parameter samples are obtained after marginalizing over the angular power spectrum instead of simply conditioning on a fiducial angular power spectrum. The prior on the CHA parameters is carefully constructed so as to maintain the positive definite nature of the covariance matrix, hence not favoring an artificially large amplitude of CHA signal.
 
We present the paper in the following manner. In section \ref{CHA_intro_and_cov} we provide an introduction to the subject of this paper, Cosmic Hemispherical Asymmetry (CHA), and describe the phenomenological model of CHA and its imprint on the covariance  matrix of the CMB temperature fluctuations. Section \ref{Prob_dist} provides an analytic insight into the probability distributions of the quantities relevant to the problem. These are the probability distributions that are sampled using the HMC algorithm. Details of the HMC sampling method are described in section \ref{Method}. In appendix \ref{Demo_on_sim_map} we demonstrate the application of the method to simulated maps. Section \ref{results} and \ref{model_comp} discuss results from the analysis of Planck-2015 SMICA temperature map and the corresponding Bayesian evidence in comparison to the statistically isotropic model respectively. Finally, we conclude  in section \ref{Conclusion}.

\section{Cosmic Hemispherical Asymmetry}\label{CHA_intro_and_cov}
\subsection{Introduction to Dipole Modulation Model}
\par CHA is modeled as dipole modulation of statistically isotropic (SI) CMB 
sky \cite{Gordon_et_el_modulation_2005}. Dipole modulated CMB temperature anisotropy field 
($\varDelta T(\hat{n}) \equiv T(\hat n)/T_0-1$)\footnote{$T_0=2.7255$ K is the all sky average temperature of CMB \cite{2009ApJ...707..916F,2014A&A...571A...9P}.}
in the direction $\hat{n}$, is given by 
\begin{equation}\label{T_tilde}
\varDelta T(\hat{n}) = (1 + A\hat{p}\cdot \hat{n}) \varDelta T^{SI}(\hat{n}).
\end{equation}
Here, $\varDelta T^{SI}(\hat{n})$ is SI CMB temperature anisotropy field. The direction of the modulation dipole is given by $\hat{p} \equiv (\theta_p, \phi_p)$ 
and $A$ is the amplitude of the dipole modulation. Within this phenomenological model, characterizing the signature of CHA then 
reduces to estimating these three parameters that capture the departure from statistical isotropy. 
Various methods have been used to study the CHA observed in the CMB. One approach is to 
construct an estimator for the amplitude and the direction of the dipole. Akrami et al. (2014) have 
used local variance estimator to look for such signal in CMB data 
\cite{Akrami-Fantaye-Shafieloo-2014}. This estimator searches for a dipole in the ``local variance 
map'' of temperature fluctuations. In Hanson et al. (2009), Quadratic Maximum Likelihood estimator 
is constructed for these parameters \cite{Hanson-Lewis}. For analysis based on BipoSH representation, the contribution of CHA is captured in $L = 1$ BipoSH coefficients 
\cite{Hajian_Souradeep_1st_BipoSH, Hajian:2004zn, Souradeep:2006dz}. Another approach is to get the likelihood distribution of these 
parameters. Hoftuft et al. (2009) sample the likelihood distribution in real space, where the likelihood is 
defined in terms of real space parameters $A$ and $(\theta_p, \phi_p)$. The likelihood is sampled using Markov 
Chain Monte Carlo method to estimate the parameters of the modulation dipole 
\cite{Hoftuft-Eriksen}. Results of these different estimators are also included in the 
\textit{Planck}-2013 \cite{Planck_2013_isotropy} and \textit{Planck}-2015 \cite{Planck_2015_isotropy} analysis. Different methods have detected the signal of CHA at a significance level around $3\sigma$ with a consistent direction.

It has been now established that the dipole modulation signal is significant only at large 
angular scales and becomes insignificant at small angular scales \cite{Hoftuft-Eriksen, Hansen:2008ym, 
Hanson-Lewis}. If we allow the amplitude of the dipole to vary with angular scale, it is called 
scale dependent dipole modulation. Given as it is written, \Eq{T_tilde} can not explicitly express the 
scale dependence of the dipole amplitude. It is best expressed in harmonic space using following equation
\begin{equation}\label{eq_T_tilde_l}
a_{lm}= a^{SI}_{lm} + \sum^{1}_{N = -1} m_{1N}(l) \int \varDelta T^{SI}(\htn) Y_{L = 1, N}
(\hat n) Y^*_{lm}(\hat n) d^2\Omega_{\hat n},
\end{equation}
where $a^{SI}_{lm}$ and $a_{lm}$ are the spherical harmonic coefficients of $\varDelta T^{SI}(\htn)$ and $\varDelta T(\htn)$, respectively, and $m_{1N}(l)$ (N = -1, 0, 1) are the spherical harmonic coefficients of the dipole field whose amplitude depends on the multipole $l$. From \Eq{eq_T_tilde_l}, it is clear that the SI temperature field $\Delta T^{SI}$ also contributes to the statistically anisotropic (non-SI) part of the signal (second term), and hence a joint estimation of the $a_{lm}$ along with the $m_{1N}$ term is required to adequately capture the signal of CHA.

Here, $m_{11}$ and  $m_{1-1}$ are complex numbers, that are related to each other by $m^{*}_{1-1} =  - m_{11}$. It is convenient to use $m_{10}$, the real part of $m_{11}$ ($m^{r}_{11}$) and the imaginary part of $m_{11}$ ($m^{i}_{11}$) as the three equivalent independent variables. These three real numbers completely determine the direction $\hat{p}$ and the amplitude $A$ of the dipole. 
The scale-dependent modulation field using real space variables amplitude $A(l)$ and the dipole direction $\hat{p}$ is expressed as
\begin{equation}\label{dipole}
 A(l) \hat p \cdot \hat n = \sum^{1}_{N = -1} m_{1N}(l) Y_{L=1,N}(\hat n).
\end{equation}
The  amplitude $A$ and the direction $(\theta_p, \phi_p)$ in terms of $m_{10}$, $m^{r}_{11}$ and $m^{i}_{11}$ are given by following equations
\begin{equation}\label{m1M_to_Athephi}
A(l) = \sqrt{\frac{3}{4 \pi}} \sqrt{m_{10}(l)^2 + 2m^r_{11}(l)^{2} + 2m^i_{11}(l)^2}, \quad
\theta_p = \cos^{-1} \Big [\frac{m_{10}(l)}{A(l)} \sqrt{\frac{3}{4 \pi}} \Big ], \quad
\phi_p = -\tan^{-1} \Big[ \frac{m^{i}_{11}(l)} {m^{r}_{11}(l) } \Big ].
\end{equation}
In most of the previous studies, a \textit{step model} of the scale dependence was assumed, wherein one assumes that the dipole modulation amplitude is nonzero with a constant value below a certain multipole $l_{cut}$ and is zero for $l > l_{cut}$. In our work, we study the \textit{power law model} for the dipole modulation amplitude $A(l)$
\begin{equation}\label{eq_power_law_Al}
 A(l) = A(l_p)\Big( \frac{l}{l_p} \Big)^{\alpha},
\end{equation}
where $A(l_p)$ is the amplitude at a chosen pivot multipole $l_p$.
The power law dependence of the asymmetry results in a power-law form harmonic space modulation field, that can be expressed as
\begin{equation}
 m_{1N}(l) = m_{1N}(l_p) \Big( \frac{l}{l_p} \Big)^{\alpha},
\end{equation}
where $m_{1N}(l_p)$ are the spherical harmonic coefficients of the dipole $A(l_p)\htp \cdot \htn$, that are related to each other by \Eq{m1M_to_Athephi}. 

The description of a non-SI CMB temperature anisotropy field that captures the power law dipole 
modulation model requires four additional parameters along with the usual spherical harmonic coefficients $a_{lm}$. In this analysis, we sample the joint probability distribution of $a_{lm}$, $C_l$, ${m_{10}, m^r_{11}, m^i_{11}}$ and $\alpha$ under the Bayesian framework using HMC method 
\cite{Duane:1987de, SantanuDas}. Previously Aiola et al. \cite{Aiola_2015_power_law} studied a power-law scale dependence form to estimate only the CHA signal. However, due to the structural form of \Eq{eq_T_tilde_l}, it is important to perform a joint analysis of both the angular power spectrum and CHA. We provide the first joint analysis in this paper accounting for all the important observational non-idealities such as partial sky coverage, anisotropic noise and instrumental beam response function.

\subsection{Covariance matrix of dipole modulated CMB temperature sky}\label{DM_model}
Since the CMB anisotropy is predicted and observed to be consistent with a Gaussian random field \cite{Planck_2015_isotropy, Planck_2013_isotropy}, its covariance matrix ($\bS$) should encode all of its information content. For an SI CMB sky, the covariance matrix in spherical harmonic space is diagonal. However, in the presence of any kind of non-SI signal (a specific form of dipole modulation is given in \Eq{eq_T_tilde_l}), the covariance matrix of $a_{lm}$ contains non-zero off-diagonal terms.  
The exact\footnote{To all orders in $m_{1N}$} covariance matrix of $a_{lm}$
for the dipole modulation case can be written as
\begin{eqnarray}\label{cov_mat}
S_{l_1m_1l_2m_2} &\equiv&  \langle a_{l_1 m_1} a^*_{l_2 m_2} \rangle = 
C_{l_1} \delta_{l_1 l_2} \delta_{m_1 m_2} \nonumber \\
&+& \frac{\Pi_{l_1l_2}}{\sqrt{ 12\pi}} 
\sum^{1}_{N = -1} (m_{1N}(l_1) C_{l_1} (-1)^{l_1 + l_2 + 1} 
+ m_{1N}(l_2) C_{l_2} ) C^{1 0}_{l_1 0 l_2 0} C^{1 N}_{l_1 m_1 l_2 m_2} \nonumber \\
&+& \frac{\Pi_{l_1l_2}}{12 \pi} \sum_{l'_1 m'_1} C_{l'_1} \Pi^2_{l'_1} 
C^{1 0}_{l_1 0 l'_1 0} C^{1 0}_{l_2 0 l'_1 0} \sum^{1}_{N_1,N_2 = -1}
m_{1N_1}(l'_1) m^{*}_{1N_2}(l'_1) C^{1 N_1}_{l_1 m_1 l'_1 m'_1} C^{1 N_2}_{l_2 m_2 l'_1 m'_1},
\end{eqnarray}
where $C_l$ is angular power spectrum of $a^{SI}_{lm}$. Here, $\Pi_{l_1l_2\hdots l_n}$ denotes $\sqrt{(2l_1 + 1)(2l_2+1)\hdots (2l_n+1)}$ and $C^{1N}_{l_{1}m_{1}l_{2}m_{2}}$ are well-known Clebsch-Gordon coefficients. 

We make use of BipoSH representation of the CMB two-point correlation function to carry out the analysis. The covariance matrix, $S_{l_1m_1l_2m_2}$, in terms of BipoSH coefficients can be expressed as \cite{Amir_Tarun_Biposh_2005}
\begin{equation}\label{cov_in_term_of_biposh}
 S_{l_1m_1l_2m_2} = (-1)^{m_2}\sum_{LN} A^{LN}_{l_1l_2} C^{LN}_{l_1m_1l_2-m_2}.
\end{equation}
The BipoSH coefficients $A^{LN}_{l_1l_2}$ act as equivalent auxiliary variables in our calculations and computations.
For $L = 0$ these capture the SI correlations and are trivially related to the angular power spectrum, $C_l$, through the relation 
\begin{equation}\label{eq_A00ll_Cl}
 A^{00}_{ll} = (-1)^l \sqrt{2l + 1} C_{l}.
\end{equation}

The leading order term of $m_{1N}$ in \Eq{cov_mat} contributes only to the off-diagonal terms of the covariance matrix. The off-diagonal elements of the covariance matrix has the 
dependence on the magnitude and direction of the modulation dipole through $m_{1N}$. The second 
order term of \Eq{cov_mat} is generally neglected on account of $m_{1N}$ being very small. However, 
this term has to be treated carefully because it leads to the modification of the diagonal of the covariance 
matrix that is used in the estimation of the angular power spectrum.\\
\textbf{Diagonal terms:}
Following is the modification to the diagonal element due to the second order term in $m_{1N}$
\begin{equation}\label{cov_mat_corr_to_diag}
\langle a_{l m} a^*_{l m} \rangle = C_{l} + \frac{\Pi^2_l}{12 \pi} 
\sum_{l' m'} C_{l'} \Pi^2_{l'} [C^{1 0}_{l 0 l' 0}]^2
\sum^{1}_{N_1,N_2 = -1} m_{1N_1}(l') m^{*}_{1N_2}(l') C^{1 N_1}_{l m l' m'} C^{1 N_2}_{l m l' m'}.
\end{equation}
Note that $\langle a_{l m} a^*_{l m} \rangle$ depends also on $m$ (and not only on $l$) through the Clebsch-Gordan coefficients. For a smooth functional form of $m_{1N}(l)$ ($m_{1N}(l+1) \approx m_{1N}(l)$), we can simplify \Eq{cov_mat_corr_to_diag} as
\begin{equation}
\tilde{C}_l = C_{l} + \frac{C_l}{4\pi} \sum^{1}_{N = -1} |m_{1N}|^2,
\end{equation}
where $\tilde C_l \equiv \langle a_{l m} a^*_{l m} \rangle$. Using \Eq{dipole}, the relative change in $C_l$ in terms of $A(l)$ can be written as
\begin{equation}\label{Delta_Cl}
 {\Delta_l \equiv \frac{\tilde{C}_l - C_{l}}{C_l} = \frac{1}{4\pi} \sum^{1}_{N = -1} |m_{1N}|^2 = \frac{A^2(l)}{3}.}
\end{equation}
In our analysis, we neglect the second order term in the covariance matrix, but provide an estimate of the relative correction in Sec. \ref{results}. It is shown to be sub-dominant and hence negligible.\\
\textbf{Off-diagonal terms:}
The $L=1$ BipoSH coefficients for the dipole modulation temperature field can be deduced from Eq. \ref{cov_mat} to be
\begin{equation}\label{DM_biposh}
 A^{1N}_{ll+1}(l_p, \alpha) = m_{1N}(l_p) G^{1}_{ll+1},
\end{equation}
where $G^{1}_{ll+1}$ is referred to as the shape factor of the specific non-SI signal. Specifically, here
\begin{equation}\label{shape_factor}
 G^{1}_{ll+1} \equiv \frac{\Pi_{ll+1}}{\sqrt{12\pi}}\Big[ \Big( \frac{l}{l_p}\Big)^{\alpha} C_{l} + 
 \Big( \frac{l+1}{l_p}\Big)^{\alpha} C_{l+1} \Big]  
C^{10}_{l0l+10}.
\end{equation}
\Eq{DM_biposh} when substituted in \Eq{cov_in_term_of_biposh} provides the off-diagonal terms of the covariance matrix in terms of $m_{1N}(l_p)$ and $G^{1}_{l l+1}$. Scale dependence is an important aspect of this empirical modeling of the CHA that is captured by the index $\alpha$ of the power law.
The shape factor encodes the signature of the dipole modulation through the mixing term between $l$ and $l\pm 1$ modes, which is sufficient to measure the non-SI signal from the map.

\section{Probability distributions}\label{Prob_dist}
In this section, we define the probability distributions of variables involved in the problem. We denote\footnote{Bold fonts are used to denote matrices.} the diagonal of $\bS$ by $\bD$ and the off-diagonal part by $\bO$ (so that $\bS = \bD + \bO$). We express the set of $a_{lm}$ by the vector $\mathbf{a}$. For the dipole modulation signal, \Eq{cov_in_term_of_biposh} and \Eq{DM_biposh} jointly express $\bO$ in terms of $C_l$, $m_{1N}(l_p)$ and $\alpha$. 

\subsection{Joint Probability distribution of \texorpdfstring{$\bS(C_l, m_{1N}, \alpha)$}{Lg} and 
\texorpdfstring{$ \{a_{lm}\}$}{Lg} }\label{3p1}
We assume that the noise in the measurement of CMB temperature is Gaussian distributed with the 
noise covariance matrix $\bN$. The probability of the data $\bd$ given $\ba \equiv \{a_{lm}\}$ is
\begin{equation}
 \calP(\textbf{d}|\textbf{a}) = \frac{1}{\sqrt{|\textbf{N}|} (2 \pi)^{n/2} } \exp \Big[ -\frac{1}{2} 
(\textbf{d} - \textbf{a})^{\dagger} \textbf{N}^{-1} (\textbf{d} - \textbf{a}) \Big ].
\end{equation}
Here, $\textbf{N}^{-1}$ is the inverse of the noise covariance matrix $\bN$ and $n$ is total number of $a_{lm}$ coefficients, $(n = l^2_{max} + 2l_{max} - 3)$, excluding monopole and dipole. 
The probability distribution of the zero mean Gaussian ${a_{lm}}$ can be written as
\begin{equation}
\calP(\textbf{a}|\textbf{S}) = \frac{1}{\sqrt{|\textbf{S}|} (2 \pi)^{n/2} } \exp \Big[ -\frac{1}{2} 
\textbf{a}^{\dagger} \textbf{S}^{-1} \textbf{a} \Big ],
\end{equation}
where $\bS^{-1}$ denotes the inverse of the covariance matrix $\bS$.
Using Bayes theorem, the joint probability distribution of $\bS$ and $\ba$ given $\bd$ in terms of 
above two probability distributions is
\begin{equation}
 \calP(\textbf{S}, \textbf{a} | \textbf{d}) = \frac{ \calP(\textbf{d} | \textbf{a}) 
\calP(\textbf{a}|\textbf{S}) \calP(\textbf{S})}{\calP(\textbf{d})}.
\end{equation}
With uniform prior $\calP(S)$ and up to a normalization constant $\calP(\textbf{d})$,
\begin{equation}\label{Prob_a_S}
 \calP(\bS, \ba | \bd) = \frac{1}{\sqrt{|\bN| |\bS| } (2 \pi)^n } \exp { \Big\{ -\frac{1}{2} 
 \Big[ (\bd - \ba)^{\dagger} \bN^{-1}  (\bd - \ba) + \ba^{\dagger} \bS^{-1} \ba \Big ] \Big\} }.
\end{equation}
In above expression, the dependence of $\calP(\bS, \ba | \bd)$ on $a_{lm}$ is explicit. $\calP(\bS, 
\ba | \bd)$ depends on angular power spectrum $C_l$, $m_{1N}(l_p)$ and $\alpha$ through $\bS$. The angular power spectrum, $C_l$, is the diagonal part of $\bS$. Dependence of $\bS$ on $m_{1N}(l_p)$ 
and $\alpha$ is given by \Eq{cov_mat}.

\subsection{Probability distribution of the dipole modulation amplitude}\label{sec_Prob_A}
We derive an analytical expression for the probability distribution of $m_{1N}$ and the dipole modulation amplitude $A$ with some approximations. 
In literature, it has been argued (on heuristic grounds) that $m_{1N}$ have Gaussian 
distribution (see section 6.3 of \cite{Planck_2015_isotropy}). As a result of this, $A$ will have 
Maxwell-Boltzmann distribution if the distribution of $m_{1N}$ is centered at the  origin. 
The probability distributions of $m_{1N}$ and $A$ play an important role in the interpretation of our results.

\par We integrate out $\ba$ from \Eq{Prob_a_S} and obtain the joint probability distribution of 
$C_l$ and $m_{1N}$. Marginalizing $\ba$ from \Eq{Prob_a_S} yields 
\begin{equation}\label{P_of_S_given_d}
 P(\bS|\bd) = \frac{1}{\sqrt{|\bN + \bS|} (2\pi)^{n/2} } 
 \exp { \Big[-\frac{1}{2} \bd^T (\bS + \bN)^{-1} \bd  \Big ] }.
\end{equation}
The explicit expression for the joint probability distribution of $C_l$ and $m_{10}$ can be written as
\begin{equation}\label{joint_P_cl_m10}
 P(C_{l}, m_{10}| \bd) = \frac{1}{\sqrt{|\bD|} (2 \pi)^{n/2} } \exp { \Big[-\frac{1}{2} 
\bd^{\dagger} \bD^{-1} \bd  \Big ]} \exp { \Big[-\frac{(m_{10} - \mu)^2}{2 \sigma^{2}} \Big]} 
\exp{\Big[\frac{\mu^{2}}{2 \sigma^{2}} \Big]},
\end{equation}
where
\begin{equation}\label{var_m10}
 \mu = \frac{\bd^{\dagger}\bD^{-1}\bO_{1}\bD^{-1}\bd }{ Tr[(\bD^{-1}\bO_{1})^{2}]}M^{-1}
\quad \text{and} \quad
 \sigma^2 = \frac{2M^{-1}}{ Tr[(\bD^{-1}\bO_{1})^{2}]},
\end{equation}
with the quantity $M$ given by
\begin{equation}
 M = \Big[ \frac{2 \bd^{\dagger} (\bD^{-1} \bO_1)^2 \bD^{-1} \bd}{Tr[(\bD^{-1}\bO_1)^2]} - 1 \Big].
\end{equation}
The detailed derivation of the \Eq{joint_P_cl_m10} is given in appendix \ref{appendix_P_Cl_m1M}. 
$M$ is a real number and is approximately equal to 1, which can be checked by replacing $M$ by its ensemble average.
The expression for $Tr[(\bD^{-1}\bO_{1})^{2}]$ is 
\begin{equation}\label{Trac_D_inv_O1_sq}
 Tr[(\bD^{-1}\bO_{1})^{2}] = \frac{1}{4 \pi} \sum^{l_{max}}_{l = l_{min}} \Big [\Big (2 + 
\frac{D_{l-1}}{D_l} + \frac{D_{l}}{D_{l-1}} \Big)l + \Big(2 + \frac{D_{l+1}}{D_l} + 
\frac{D_{l}}{D_{l+1}} \Big)(l + 1) \Big],
\end{equation}
where $D_l = C_l + N_l$ and $(l_{min}, l_{max})$ is the range of multipoles that are dipole 
modulated. Inspection of \Eq{joint_P_cl_m10} reveals that $m_{10}$ is Gaussian distributed with 
mean $\mu$ and variance $\sigma^2$.  It is informative to have a qualitative look at the expression for $\sigma^2$. As $C_l \approx C_{l+1}$ and CMB measurement are noise sub-dominant at large angular scales ($N_l << C_l$), the variance $\sigma^2$ is approximately equal to $2 \pi/(l^2_{max} + 2l_{max} - l^2_{min} + 1)$. This implies (as expected) that the variance is inversely proportional to the total number of independent modulated modes. This compact form is obtained for the scale independent $m_{1N}$. However, the same essence is also preserved for any scale dependent modulation field.

The components $m_{10}$ and $m_{11}$ have same variance, hence $m^r_{11}$ and $m^i_{11}$ have $\sigma^2/2$ as 
their variance. For the purpose of discussing the probability distribution, it is convenient to deal with the variables that have the same variance. So, we define 
three variables $w_x$, $w_y$, and $w_z$ as 
\begin{equation}\label{def_wxyz}
 w_z = m_{10},\, w_x = -\sqrt{2}m^r_{11},\, \text{and} w_y = \sqrt{2} m^i_{11}.
\end{equation}
Then, $w_x, w_y, w_z$ are distributed with same variance $\sigma^2$. The norm of the signal $r$ can be defined as 
\begin{equation}\label{def_r}
r \equiv \sqrt {w^2_z + w^2_x + w^2_y } = \sqrt {m_{10}^{2} + 2{m^r_{11}}^{2} + 2{m^i_{11}}^{2} }.
\end{equation}
For a SI CMB map, $w_x$, $w_y$, and $w_z$ are Gaussian distributed with zero mean. Hence $r$ is distributed 
according to Maxwell-Boltzmann distribution
\begin{equation}\label{Prob_r_SI}
 \calf(r) = \sqrt{\frac{1}{(2 \pi \sigma^2)^3} } 4 \pi r^2 
 \exp { \Big[-\frac{r^2}{2 \sigma^2} \Big ]},
\end{equation}
where $\sigma^2$ is given by \Eq{var_m10}. The relation $r = \sqrt{\frac{4 \pi}{3}} A$ leads us to 
the probability distribution of the dipole amplitude $A$ in case of SI map as 
\begin{equation}\label{MB_dist_A_SI}
 \calf(A) = \frac{8\pi}{3} \sqrt{\frac{2}{3}} \frac{A^2}{\sigma^3} 
 \exp { \Big[-\frac{2 \pi A^2}{3 \sigma^2} \Big ]}.
\end{equation}
For the dipole modulated CMB sky, $w_x$, $w_y$, and $w_z$ are Gaussian distributed but are not centred at 
origin. We denote the Maximum Likelihood values of $w_x$, $w_y$, and $w_z$ by $w_{x*}$, $w_{y*}$, and $w_{z*}$ respectively, and the corresponding $r_*$ denotes the signal amplitude through \Eq{def_r}.
Then the distribution of $r$ marginalized over $(\theta, \phi)$ is
\begin{equation}\label{Prob_r_DM}
 \calf(r) = \frac{r}{\sigma^2} \sqrt{\frac{r}{r_*}} 
 \exp \Big[ -\frac{r^2 + r^2_*} {2 \sigma^2} \Big] I_{1/2}(\frac{r r_*}{\sigma^2}),
\end{equation}
where $I_{1/2}(x)$ is a modified Bessel function of first kind of order $1/2$,
\begin{equation}
 I_{1/2}(x)=\sqrt{\frac{x}{2\pi}} \int^{\pi}_0 d\theta \sin(\theta) \exp \Big[ x \cos(\theta) \Big].
\end{equation}
The probability distribution for $A$ is
\begin{equation}\label{Prob_A_DM}
  \calf(A) = \frac{4 \pi}{3} \frac{A}{\sigma^2} \sqrt{\frac{A}{A_* }} 
  \exp \Big[ -\frac{2\pi(A^2 + A^2_* )}{3 \sigma^2} \Big] I_{1/2}(\frac{4 \pi A A_* }{3\sigma^2}),
\end{equation}
where $A_*$ is the dipole amplitude corresponding to the radial coordinate $r_*$. For an SI map, $A_*$ and $r_*$ are equal to zero. It is important to note that $A_*$ is not same as the maximum probable value of $A$ where $\calf(A)$ attains its maximum.

\section{Method: Hamiltonian Monte Carlo sampling for non-SI CMB sky in presence of observational non-idealities}\label{Method}
For the given problem, we deal with $\approx l^2_{max} + 3 l_{max}$ parameters. Markov Chain Monte Carlo method with the Metropolis-Hastings algorithm is known to have a much smaller acceptance rate in high dimensional parameter space; hence it is not practical for our task. This problem is circumvented in the Gibbs sampling method, which has been used in the CMB analysis for the estimation of the angular power spectrum \cite{Wandelt:2003uk, Eriksen:2004ss} as well as component separation \cite{Eriksen:2007mx}. The Gibbs sampling uses the conditional distributions of the parameters as proposal distributions. However, in our problem, with the addition of dipole modulation parameters, it is not straightforward to get the conditional distributions and draw samples from those distributions. In comparison, HMC is an efficient method to draw samples from high dimensional distributions. It makes use of Hamiltonian dynamics to propose the next sample, as discussed next in this section.

In this section, we describe  the Hamiltonian Monte Carlo (HMC) method 
employed to sample the posterior probability distribution. We give details of the analytical computations required to implement this Monte Carlo method.  The mask, anisotropic noise, and the beam need to be accounted for to analyze the Planck CMB maps. The validation of this method is performed on simulated non-SI maps obtained using CoNIGS \cite{Mukherjee_conigs} as described in Appendix \ref{Demo_on_sim_map}.

\subsection{Review of the basic formalism as a HMC sampling problem}
HMC has been applied to various problems in cosmology, notably for 
cosmological parameter estimation in \cite{Hajian_HMC}, for CMB power spectrum estimation in 
\cite{Taylor_Cl_with_HMC_2008}, for estimation of the large scale structure power spectrum in \cite{Jasche_HMC_LSS, Jasche_Wandelt2013}, the inference of non-linear dynamics of large scale structure \cite{2013MNRAS.432..894J,2013ApJ...772...63W}, and lensing potential reconstruction from lensed CMB in \cite{Anderes_Wandelt_Lavaux_bayes_lensing_2015}. HMC makes use of Hamiltonian Dynamics to sample 
a given probability distribution $\calP(q)$ of random variable $q$ using a Hamiltonian 
($\calH$) defined in the following way 
\begin{equation}\label{test_H}
 \calH = \frac{p^2}{2\mu} - \ln (\calP(q)),
\end{equation}
where $q$ represents the position of this particle and $p$ is the conjugate momentum corresponding 
to $q$. The quantity $- \ln (\calP(q))$ acts as the potential energy and $\mu$ is ``the mass'' 
assigned to ``the particle''. In HMC, the choice of the mass parameter, $\mu$ is specific to the problem (akin to the width of the proposal distribution in Markov Chain Monte Carlo method). Hamiltonian 
dynamics stipulates the time evolution of $q$ and $p$ through the Hamilton's equations
\begin{equation}
 \dot{q} \equiv \frac{dq}{dt} = \frac{\pd \calH}{\pd p}  \quad \text{and} \quad
 \dot{p} \equiv \frac{dp}{dt} = -\frac{\pd \calH}{\pd q}.
\end{equation}
For the detailed steps of HMC we follow the algorithm given in \cite{Hajian_HMC} and to 
evolve the sympletic Hamiltonian dynamics equations we use the Forest-Ruth algorithm (see \cite{SantanuDas} for details of the 
implementation). HMC algorithm generates the samples of $p$ and $q$ drawn from the following probability distribution
\begin{equation}
 \exp(-\calH) = \exp\Big[ -\frac{p^2}{2\mu} \Big] \calP(q).
\end{equation}
In the end, we marginalize over $p$ and obtain a fair sample of $q$ drawn from $\calP(q)$.

\par In our problem, the variables of dynamics are the parameters ($a^r_{lm}$, $a^i_{lm}$, $C_l$,
$m_{10}$, $m^r_{11}$, $m^i_{11}$, $\alpha$) and corresponding momentum ($p^r_{lm}$, $p^i_{lm}$, 
$p_{C_l}$, $\bar{p}_{10}$, $\bar{p^r}_{11}$, $\bar{p^i}_{11}$, $p_{\alpha}$).
For the probability distribution defined in \Eq{Prob_a_S}
\begin{eqnarray}\label{def_H}
 \calH(\bp_{lm}, \bp_{C_l}, \bar{\bp}_{1N}, p_{\alpha}, \ba^r, \ba^i, \{C_l\}, \textbf{m}, \alpha) 
&=& \sum^{l_{max}}_{l = 2} \Big\{ \frac{p^2_{C_l}}{2\mu_{C_l}} + \sum^{l}_{m = 0} 
\Big[\frac{{p^r}^{2}_{lm}}{2\mu^r_{a_{lm}}} + \frac{{p^i}^{2}_{lm}}{2\mu^i_{a_{lm}}} \Big] \Big\} + 
\frac{p^2_{\alpha}}{2\mu_{\alpha}} \\ \nonumber
&+& \frac{\bar{p}^2_{10}}{2\bar{\mu}_{10}} + \frac{ {\bar{p^r}}^2_{11} }{2\bar{\mu}^r_{11}} + 
\frac{ {\bar{p^i}_{11}}^2 }{2 \bar{\mu}^i_{11} } 
-\ln \Big [ \calP \Big( S_{lml'm'}, a_{lm} \Big | d_{lm} \Big) \Big].
\end{eqnarray}
In above equation, $\textbf{p}_{lm}$ and $\bp_{C_l}$ stand for the set of conjugate momenta for 
$a_{lm}$ and $C_l$ respectively. $\bar{\textbf{p}}_{1N}$ stands for conjugate momenta 
$\{ \bar{p}_{10}, \bar{p}^r_{11}, \bar{p}^i_{11}\}$ of ($m_{10}, m^r_{11}, m^i_{11} $) and 
$p_{\alpha}$ is conjugate momentum of $\alpha$. Further, $\mu^r_{a_{lm}}, \mu^i_{a_{lm}}, \mu_{C_l}$ are the 
masses assigned to the real part of $a_{lm}$, the imaginary part of $a_{lm}$ and $C_l$, respectively. Whereas, $\bar{\mu}_{10}$, $\bar{\mu}^r_{11}$, $\bar{\mu}^i_{11}$, and $\mu_{\alpha}$ are the masses assigned to $m_{10}$, $m^r_{11}$, $m^i_{11}$, and $\alpha$, respectively.
We follow arguments given in \cite{SantanuDas} and \cite{Taylor_Cl_with_HMC_2008} to choose masses 
for $a^r_{lm}$, $a^i_{lm}$, and $C_l$. The masses for $a^r_{lm}$ and $a^i_{lm}$ are chosen equal to inverse of their variance. Therefore,
\begin{equation}
 \mu^r_{a_{lm}} = \mu^i_{a_{lm}} = (2/C_l + 2/N_l) \quad \text{for} \quad m \neq 0 
 \quad \text{and} \quad \mu_{a_{l0}} = (1/C_l + 1/N_l).
\end{equation}
The mass for $C_l$ is chosen as inverse of the cosmic-variance of $C_l$
\begin{equation}
 \mu_{C_l} = \frac{2l + 1}{2C^2_l}.
\end{equation}
In the implementation of the computation, we find it convenient to use scaled parameters $10^2 m_{1N}$ to ensure that variable takes values of
the order unity. We choose unit masses corresponding to $m_{10}, m^r_{11}, m^i_{11}$, and $\alpha$.

To perform the Hamiltonian dynamics, it is necessary to compute derivatives of the conjugate momentum. This involves taking derivative of the distribution function with respect to the corresponding parameter. For the probability distribution given in \Eq{Prob_a_S}, it is possible to analytically 
compute the derivatives with some approximations, bypassing the need for numerical computation of 
the derivatives. The detailed expressions for the momentum derivatives for all the parameters are given in Appendix \ref{Appendix_PALMll_dot}. The formalism applies to any covariance matrix with non-zero off-diagonal terms. Hence, we discuss the methodology for the general modulation case and note explicitly if any result is specifically applicable only to  the scenario of dipole modulation.

We simulate the Hamiltonian dynamics for the Hamiltonian given in \Eq{def_H} using above equations. 
Following the HMC algorithm then leads to the samples of ($a^r_{lm},a^i_{lm},C_l,m_{10},m^r_{11}, 
m^i_{11}, \alpha$) which are drawn from the multidimensional probability distribution 
$\calP \Big(S_{lml'm'}, a_{lm} \Big | d_{lm} \Big)$, where $S_{lml'm'}$ depends on $(C_l, 
m_{10}, m^r_{11}, m^i_{11}, \alpha)$.

\subsection{Incorporation of observational non-idealities}
The observed sky temperature contains the true signal marred by several observational non-idealities the dominant effects being the residual foreground contamination, anisotropic noise, and instrumental beam.
CMB data is masked to reduce the residual foreground contaminations of other emissions from the galactic plane and extragalactic point sources. Hence we deal with a partial CMB sky appropriately masked to minimize residual foreground contamination for the purpose of data analysis. To deal with the masked sky and anisotropic noise, we resort to the real space variables.

Let $\bT$ represent the vector of CMB temperature 
signal in the pixel space and $\bd$ the data vector in the same space. We assume that noise is 
Gaussian distributed with variance $\sigma^2_i$ for the $i^{th}$ pixel and the noise between two 
pixels are not correlated. Then the probability distribution of $\bT$ given $\bd$ is
\begin{equation}
 \calP(\bT | \bd) = \frac{1}{(2 \pi)^{N_{pix}/2} |\bN|^{1/2}} 
 \exp { \Big[ -\frac{1}{2} \sum^{N_{pix}}_{i = 1} \frac{(d_i - T_i)^{2}}{\sigma^{2}_{i} } \Big]},
\end{equation}
where $N_{pix}$ is the total number of the pixels. We model the masked portion by setting the noise variance in those pixels to a large (effectively infinite) value. 
We take into account the finite resolution of CMB data that is convolved with the beam through a circularly symmetric beam transfer function $b_{l}$. Hence, the time derivative of conjugate momentum corresponding to $a_{lm}$ is obtained as
\begin{equation}\label{plm_mask}
\dot{p}_{lm} = -\frac{1}{2} \sum_{l_1 m_1} S^{-1}_{l_1m_1lm} a^{*}_{l_1m_1} 
+ \frac{1}{2} \sum^{N_{pix}}_{i = 1} \frac{1}{\sigma^2_i} \Big[ \sum_{l'm'} 
(d_{l'm'} - b_l p_l a_{l'm'}) Y_{l'm'}(\hat{n_i}) \Big]Y_{lm}(\hat{n_i}),
\end{equation}
where $p_l$ is the pixel window function in 
harmonic space. Note, the anisotropic noise is implicitly incorporated in the above equation due to the dependence of the $\sigma^2_i$ on pixels.
We used \texttt{map2alm} and \texttt{alm2map} routines of \texttt{HEALPix}\footnote{\url{http://healpix.sourceforge.net}}  \cite{Gorski_healpix_paper}
to compute the second term in the above equation. Given the difference map in harmonic space, $(d_{l'm'} - b_l p_l 
a_{l'm'})$, the \texttt{alm2map} subroutine executes the following computation
\begin{equation}\label{eq_diff_map}
 M_D(\hat n_i) = \sum_{l'm'} \Big[ (d_{l'm'} - b_l p_l a_{l'm'}) Y_{l'm'}(\hat{n_i}) \Big].
\end{equation}
With the map $M_D(\hat n_i)$ as input to the subroutine \texttt{map2alm}, we get following quantity as the output
\begin{equation}
\sum^{N_{pix}}_{i = 1} \frac{M_D(\hat n_i)}{\sigma^2_i} Y_{lm}(\hat{n_i}) \frac{4 \pi}{N_{pix}}.
\end{equation}
This section concludes all the essential ingredients to perform an HMC analysis of a non-SI CMB sky in the presence of the anisotropic noise and mask.

\section{Results: Analysis of Planck-2015 CMB temperature map}\label{results}
\begin{figure}[t]
\centering
\subfigure[ ]{\label{fig_map1_a}  
\includegraphics[width=0.65\linewidth]{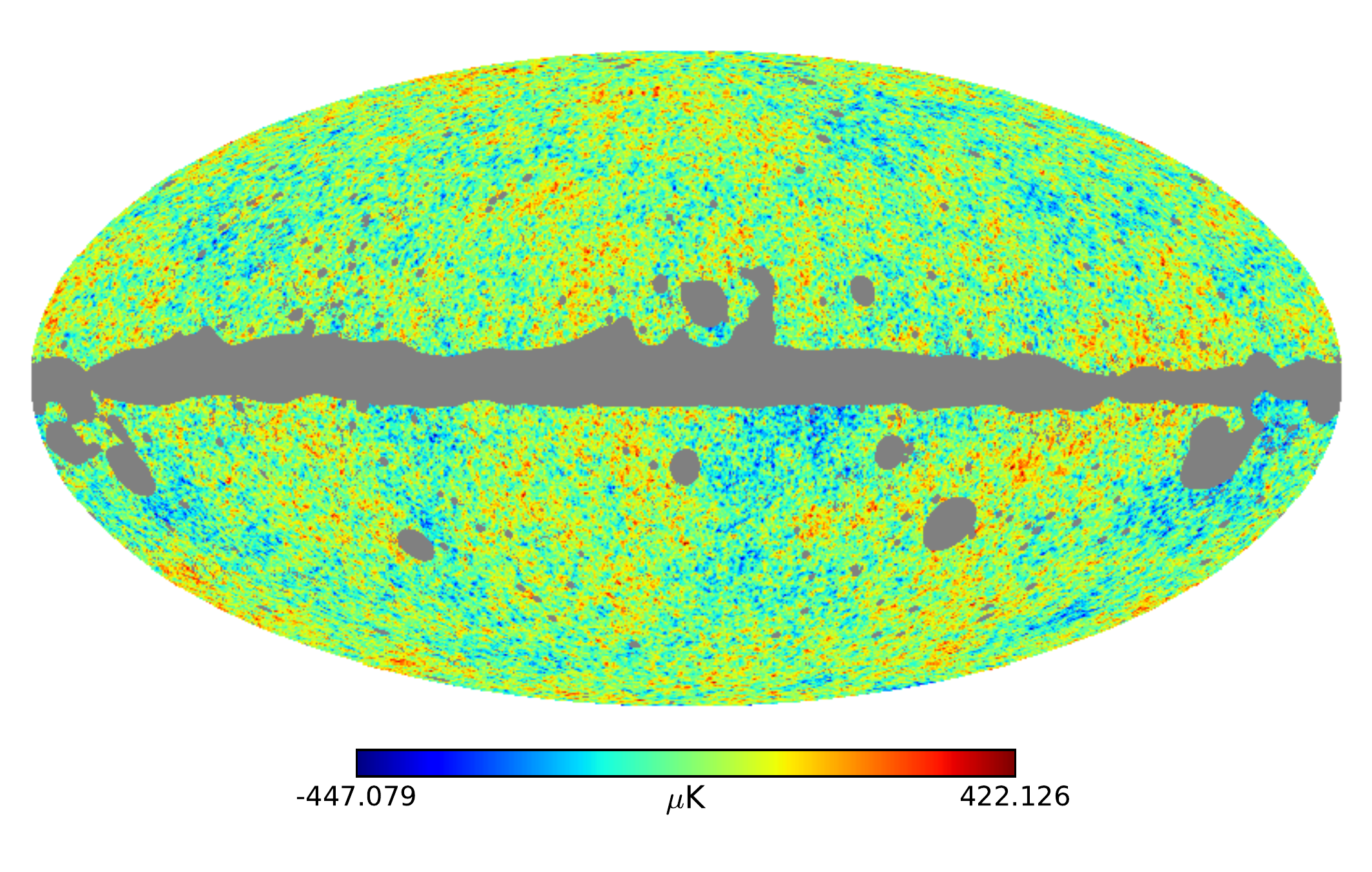}}
\subfigure[ ]{\label{fig_map1_c} 
\includegraphics[width=0.65\linewidth]{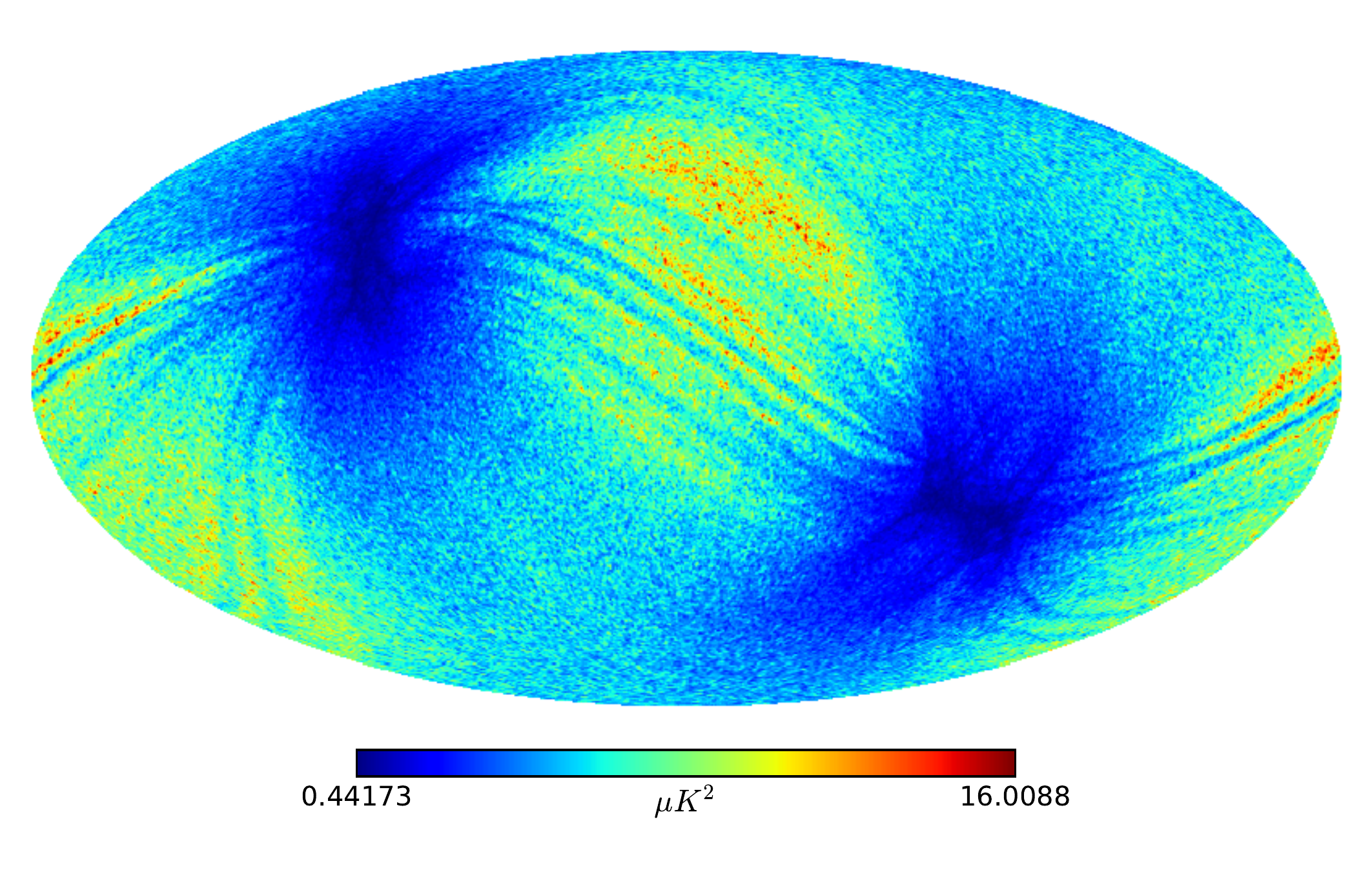}}
\caption{Figure \ref{fig_map1_a} shows SMICA temperature map (\texttt{NSIDE = 1024}) with $16\%$ masked region shown in grey. Figure \ref{fig_map1_c} shows the noise variance map (\texttt{NSIDE = 256}) used in the analysis, obtained using 100 FFP8.1 noise simulations}\label{fig_map1}
\end{figure}

In this section, we provide the result of the analysis done on the CMB temperature anisotropy map 
provided by \textit{Planck} collaboration\footnote{Planck-2015 temperature data are substantially similar to the Planck-2018 release that focused on large angle polarization anisotropies}. We use the SMICA CMB temperature map\footnote{Name of the 
file:\texttt{COM\_CMB\_IQU-smica\_1024\_R2.02\_full.fits} available at \url{https://pla.esac.esa.int/pla/\#maps}} and the corresponding SMICA mask with sky fraction $84\%$ in this analysis. The masked SMICA map is depicted in the figure \ref{fig_map1_a}.
Since we are only interested in the large angular scale anomalies, we choose to carry out our analysis at the resolution of \texttt{NISDE} = 
256 to reduce the computational cost. The estimation of the noise variance in every pixel is obtained from $100$ Full Focal Plane (FFP8.1) noise simulations \cite{Planck_FFP_2015} available at NERSC\footnote{
http://crd.lbl.gov/departments/computational-science/c3/c3-research/cosmic-microwave-background/cmb-data-at-nersc/} which we first low pass filter at multipole l = 256 before computing noise variance. The noise variance map obtained in this way and used in our analysis is shown in figure \ref{fig_map1_c}.
\label{fig_maps}

We performed the analysis assuming both the models for dipole modulation: (1) step model and (2) 
power law model. The analysis with the step model is useful for comparison of our results with the previous studies \cite{Planck_2015_isotropy,Planck_2013_isotropy}. Detailed results of the analysis with step model are given in section \ref{step_model_SMICA}. Analysis for the power law model is given in section \ref{powerlaw}. A summary of the results for both these cases is provided in Table \ref{tab:title2} and \ref{tab:title3}.
We give the mean values and standard deviation of the parameters $m_{10}$, $m^r_{11}$, and $m^i_{11}$. Values of $A_{*}$, $l$, and $b$ given in the table are obtained from the mean values of $m_{10}$, $m^r_{11}$, and $m^i_{11}$, whereas their standard deviations are obtained from their respective Monte Carlo samples.

\begin{table}[t] 
\centering
\caption{Analysis details and summaries of parameter posterior marginals (Step model)}
\label{tab:title2} 
\begin{tabular}{ |m{1.8cm}||m{1.5cm}|m{1.5cm}| m{1.5cm}| m{1.5cm}| m{1.0cm}| m{1.0cm}|}
\hline
\multicolumn{7}{|c|}{ Details of the map used in analysis: \texttt{NSIDE} = 256, $f_{sky} = 84\%$, $l_{cut} = 64$ } \\
\hline
 Parameter & $m_{10}$ & $m^r_{11} $ & $m^i_{11}$ & $A_*$ & $l$ & $b$ \\
 \hline
 Mean      & -0.039   & 0.035  & -0.056  & 0.050 & $238.0^o$ & $-22.4^o$\\
\hline
 Standard Deviation & 0.038 & 0.029  & 0.029  & 0.018 & 30.4 & 20.6\\
\hline
\end{tabular}
\end{table}

\begin{table}[t]
\centering
\caption{Analysis details and summaries of parameter posterior marginals (Power law model)}\label{tab:title3}
\begin{tabular}{ |m{1.8cm}||m{2.0cm}|m{1.5cm}| m{1.5cm}| m{1.0cm}|  m{1.0cm}| m{1.0cm}| m{1.0cm}|}
\hline
\multicolumn{8}{|c|}{Prior on $A(l_p)$ is chosen such that $A(l) < 1/2 \quad \forall \quad l$ (see section \ref{model_comp} for details)} \\
\hline
\multicolumn{8}{|c|}{Prior on $\alpha$: Uniform prior over the range $\alpha = -2$ to $\alpha = 0$}\\
\hline
\multicolumn{8}{|c|}{Details of the map used in analysis: \texttt{NSIDE} = 256, $l_{max} = 256$, $f_{sky} = 84\%$} \\
\hline
 Parameter & $m_{10}(l_p = 16)$ & $m^r_{11}(l_p) $ & $m^i_{11}(l_p) $ & $\alpha$ & $A_*(l_p)$ & $l$ & $b$ \\
 \hline
 Mean     & -0.044   & 0.033   & -0.081  & -0.92 & 0.064 & $ 247.8^o$ & $ -19.6^o$\\
\hline
 Standard Deviation & 0.043 & 0.031  & 0.032   & 0.22 & $0.022$ & 22.0 &  18.0\\
\hline
\end{tabular}
\end{table}

\subsection{Step-model of scale dependence}\label{step_model_SMICA}
\begin{figure}[H]
\includegraphics[width=0.8\linewidth,center]{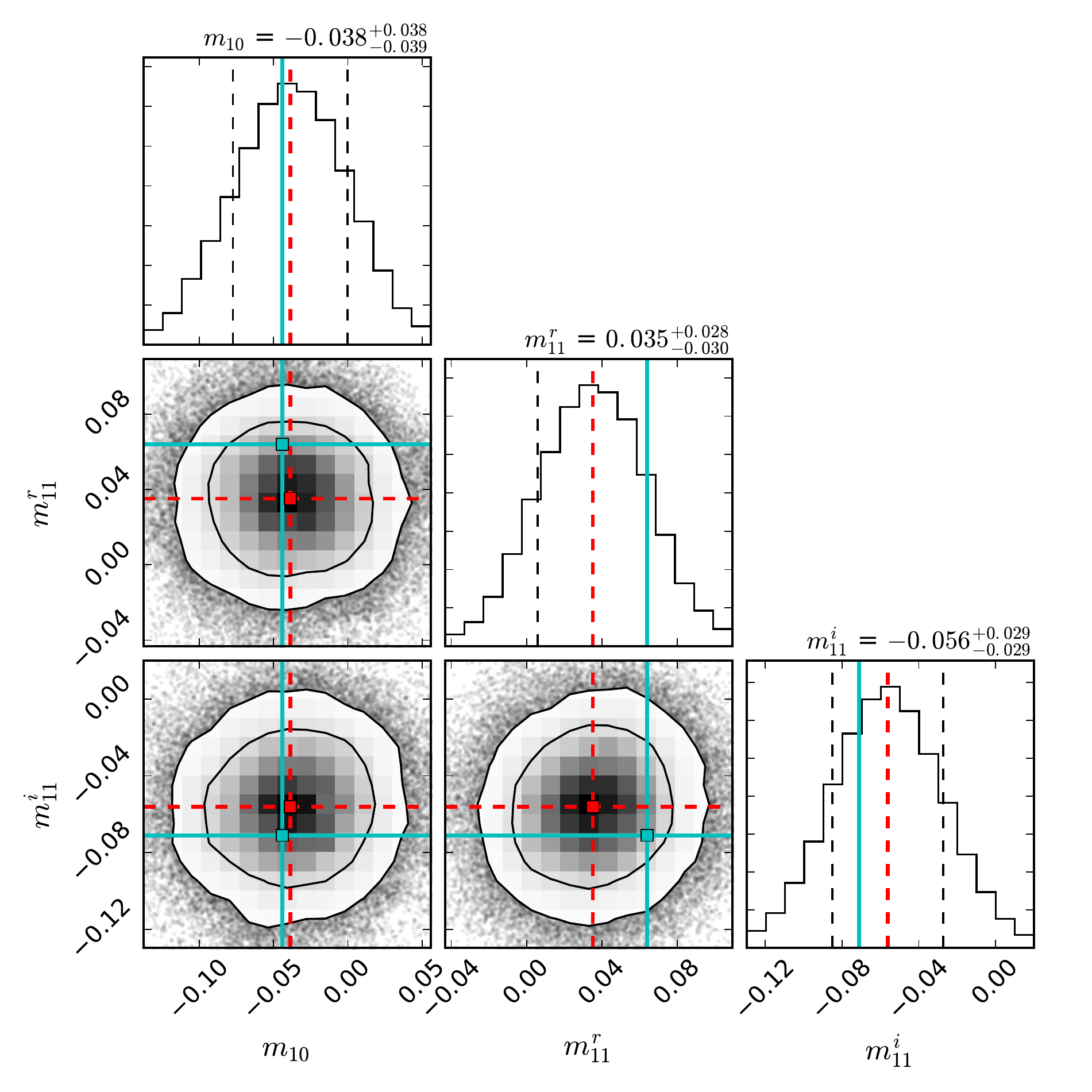}
\caption{The plot shows the joint and marginalized posterior distributions of parameters 
$m_{10}$, $m^{r}_{11}$, and $m^{i}_{11}$ for the Planck SMICA map. In 2D distributions, contours show regions of distribution containing 68\% and 90\% samples. In 1D distributions, black dashed 
lines mark 16 and 84 percentiles of the distribution. These percentile values and the median value are given in the title above the histogram of respective parameter. Red dashed lines mark the mean of the distribution. Respective mean and standard deviation of the 
parameters ($m_{10}, m^{r}_{11}, m^{i}_{11}$) are $(-0.039, 0.038)$, $(0.035, 0.029)$, 
$(-0.056, 0.029)$. Cyan lines mark the BipoSH minimum variance estimate of the corresponding parameters from Ref. \cite{Planck_2015_isotropy}. }
\label{fig_smica_m1M_step_model}
\end{figure}

To compare the results of our method with the published literature \cite{Planck_2013_isotropy, Planck_2015_isotropy},
we present the results of our analysis with a step model. In the step model  the  dipole 
modulation parameters take non-zero, constant values  over the multipole range $l = 2$ to $l = l_{cut}$. For the
results presented in this section we take $l_{cut} = 64$, a choice motivated by the literature 
\cite{Planck_2013_isotropy, Planck_2015_isotropy}. Figure \ref{fig_smica_m1M_step_model} shows 
the posterior distribution of $m_{10}$, $m^{r}_{11}$, and $m^{i}_{11}$ for SMICA map obtained from 
$10^5$ Monte Carlo samples after removing the $5 \times 10^4$ samples as Burn-In. 
The inferred joint distribution of $m_{10}, m^r_{11}$, and $m^i_{11}$ parameters do not show any significant correlation among these variables as evident from the joint distributions of these parameters given in figure \ref{fig_smica_m1M_step_model}. 
We compare our estimates of the dipole parameters with the estimates of corresponding parameters obtained using the minimum variance estimation method for BipoSH (BipoSH-MVE) given in 
\cite{Planck_2015_isotropy}. We note that the current estimate of the parameter $m^r_{11}$ deviates from the BipoSH-MVE estimate by around 
$1 \sigma$. Other two parameters agree with their corresponding value from BipoSH-MVE.
It is evident from the mean ($\mu$) and standard deviation ($\sigma$) of these parameters that each 
parameter is away from zero more than one $\sigma$. In the analysis, assumption about the multipole range 
of modulation goes in the form of shape factor. For the range of multipoles considered here $l = 
2$ to $l = 64$, $\sigma = 0.039$. The standard deviation for $m^r_{11}$ and $m^i_{11}$ is 
$\sigma/\sqrt{2} = 0.027$. The standard deviations of sampled distributions are close to the values 
expected from the analytical arguments. 

\begin{figure}[H]
\centering
\subfigure[ ]{\label{fig_smica_A_step_model}
\includegraphics[width=0.7\textwidth]{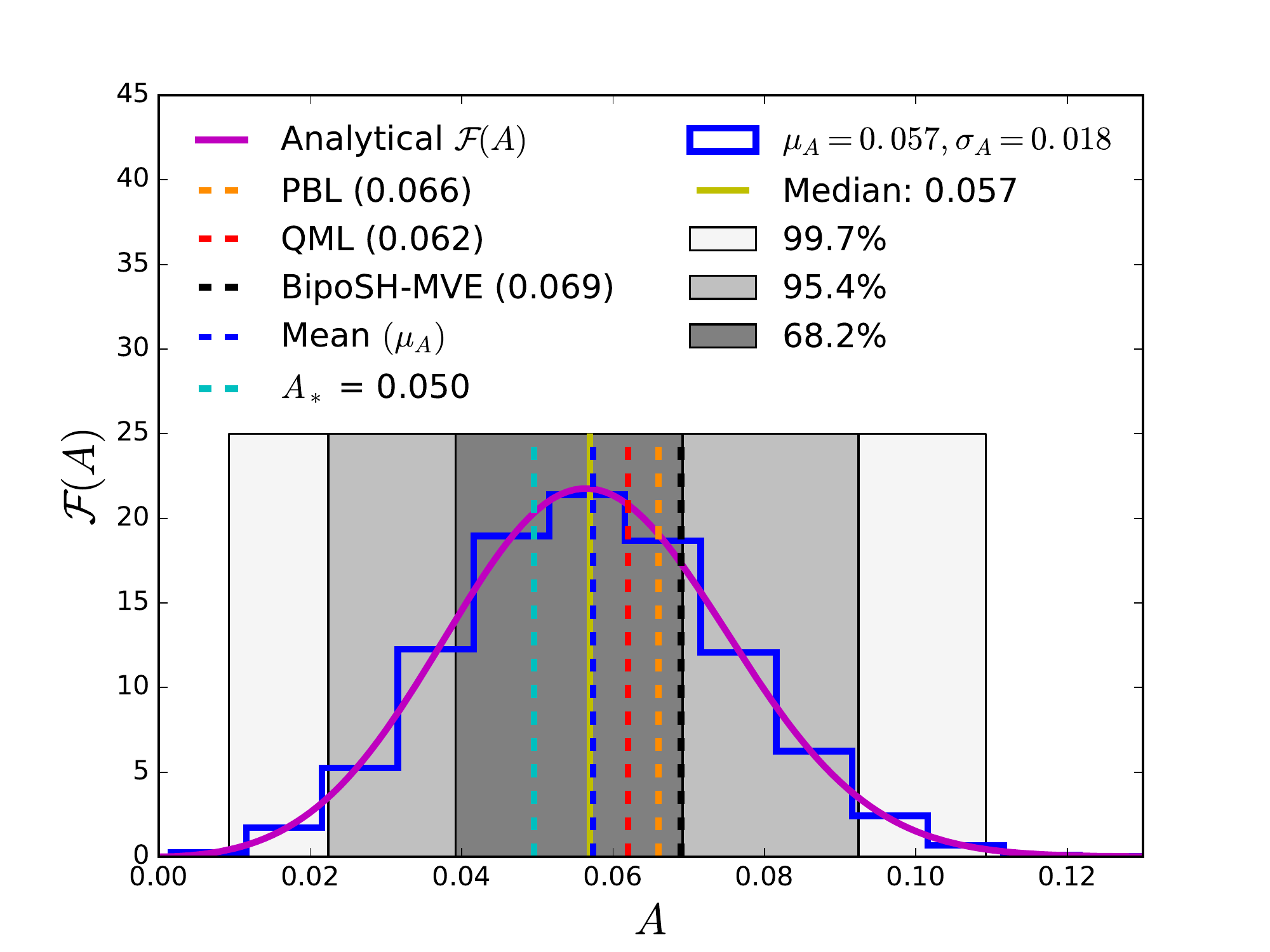}}
\subfigure[ ]{\label{fig_smica_theta_phi_step_model}
\includegraphics[width=0.7\textwidth]{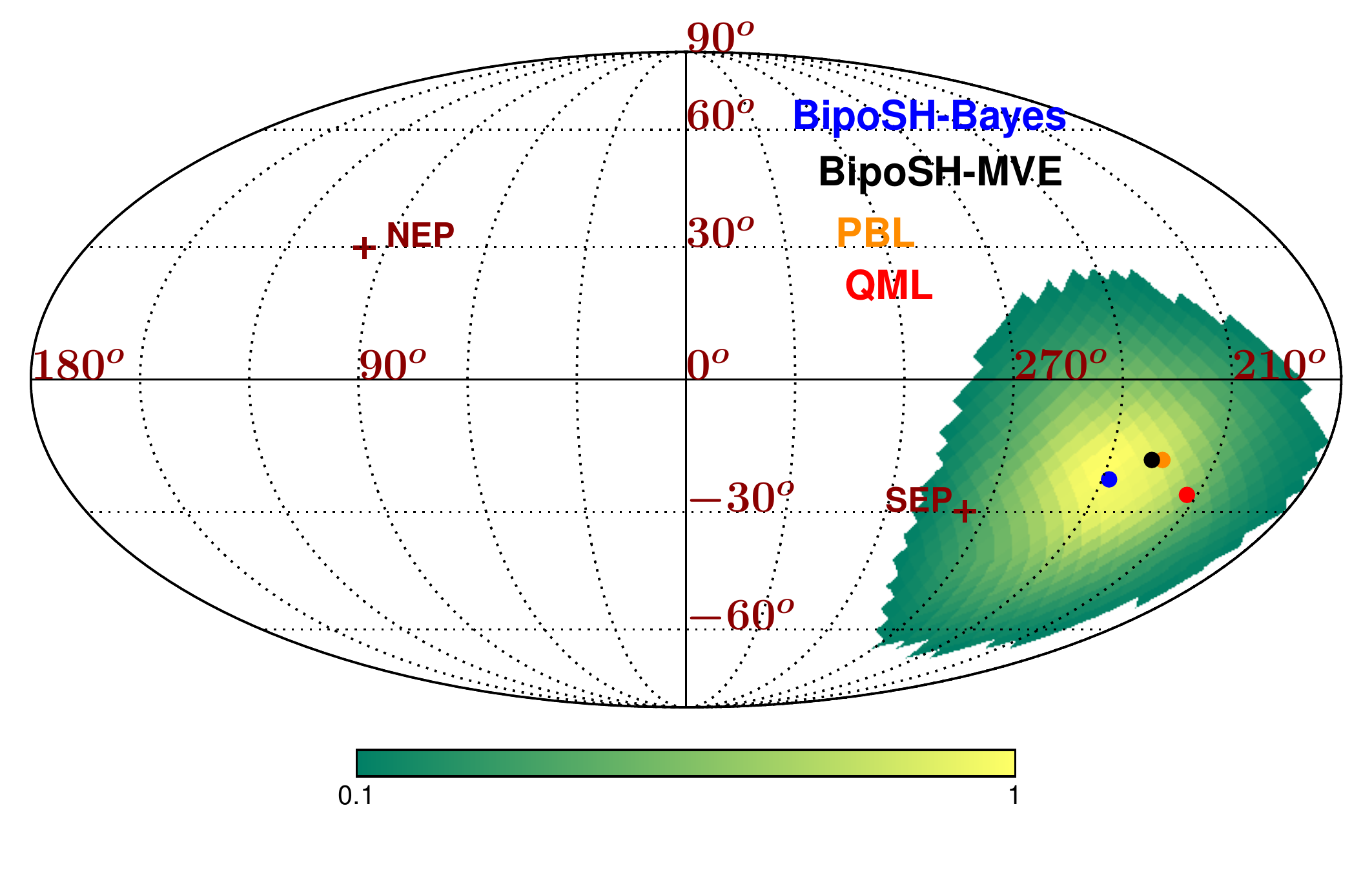}}
\caption{\textbf{(a)} The histogram in blue shows the normalized  posterior distribution of A. Shaded regions mark the 68\%, 95\% and 99.7\% areas under the histogram. Dashed blue vertical line marks the mean $(\mu_{A})$ of the distribution, $\sigma_{A}$ is the standard deviation of the distribution. Along with the mean, we also show estimates of $A$ given in \cite{Planck_2015_isotropy}: (1) Pixel Based Likelihood (PBL) estimate (orange  dashed) (2) Quadratic Maximum Likelihood (QML) estimate (red dashed),  and (3) BipoSH Minimum variance estimate (BipoSH-MVE) (black dashed). Cyan dashed line marks $A_*$, the dipole amplitude corresponding to the mean values of $m_{10}, m^r_{11}, m^i_{11}$. The magenta curve shows the analytical distribution of $A$ given in \Eq{Prob_A_DM} with $ A_* = 0.050$. \textbf{(b)} The posterior distribution of $\theta_p$ and $\phi_p$ for SMICA map. $(\theta_p, \phi_p)$ values are binned using \texttt{HEALPix} \texttt{NSIDE} = 16 grid. The resultant histogram is normalized with respect to its peak value and is further smoothed by a Gaussian with standard deviation 3.7 degrees for presentation purpose. We adopt the galactic coordinate system for this plot. 
Blue dot represents the maximum a posteriori value of the dipole direction (which corresponds to the direction obtained from the mean values of $m_{10}, m^r_{11}, m^i_{11}$) called BipoSH-Bayes and has galactic coordinates $(l, b) = (238.0^o, -22.4^o)$. Also shown are the estimates from \cite{Planck_2015_isotropy}: (1) PBL estimate (orange) $(l, b) = (225^o, -18^o)$, (2) QML estimate (red) $(l, b) = (213^o, -26^o)$, and (3) BipoSH-MVE (black) $(l, b) = (228^o, -18^o)$.}
\end{figure}

The combined effect of this deviation from 
the origin, on the dipole amplitude $A$, is reflected in the posterior distribution of $A$ given in the 
figure \ref{fig_smica_A_step_model}. The posterior distributions of $A,$ in 
figure \ref{fig_smica_A_step_model} and the posterior distribution of $\theta_p, \phi_p$ in figure 
\ref{fig_smica_theta_phi_step_model} are obtained by transforming the Monte Carlo chains of 
\{$m_{10}$, $m^{r}_{11}, m^{i}_{11}$\} to \{$A, \theta_p, \phi_p$\} using coordinate transformation equations 
given in \Eq{m1M_to_Athephi}. We also quote the estimates of \{${A, \theta_p, \phi_p}$\} from the previous analysis \cite{Planck_2015_isotropy} in figure \ref{fig_smica_A_step_model} and \ref{fig_smica_theta_phi_step_model}. This analysis shows good agreement with all the previous analysis, with the maximum difference (about  {one} $\sigma$) with the BipoSH minimum variance estimator.

\subsection{Power-law model of scale dependence}\label{powerlaw}
\begin{figure}[t]
\includegraphics[width=.8\linewidth, 
center]{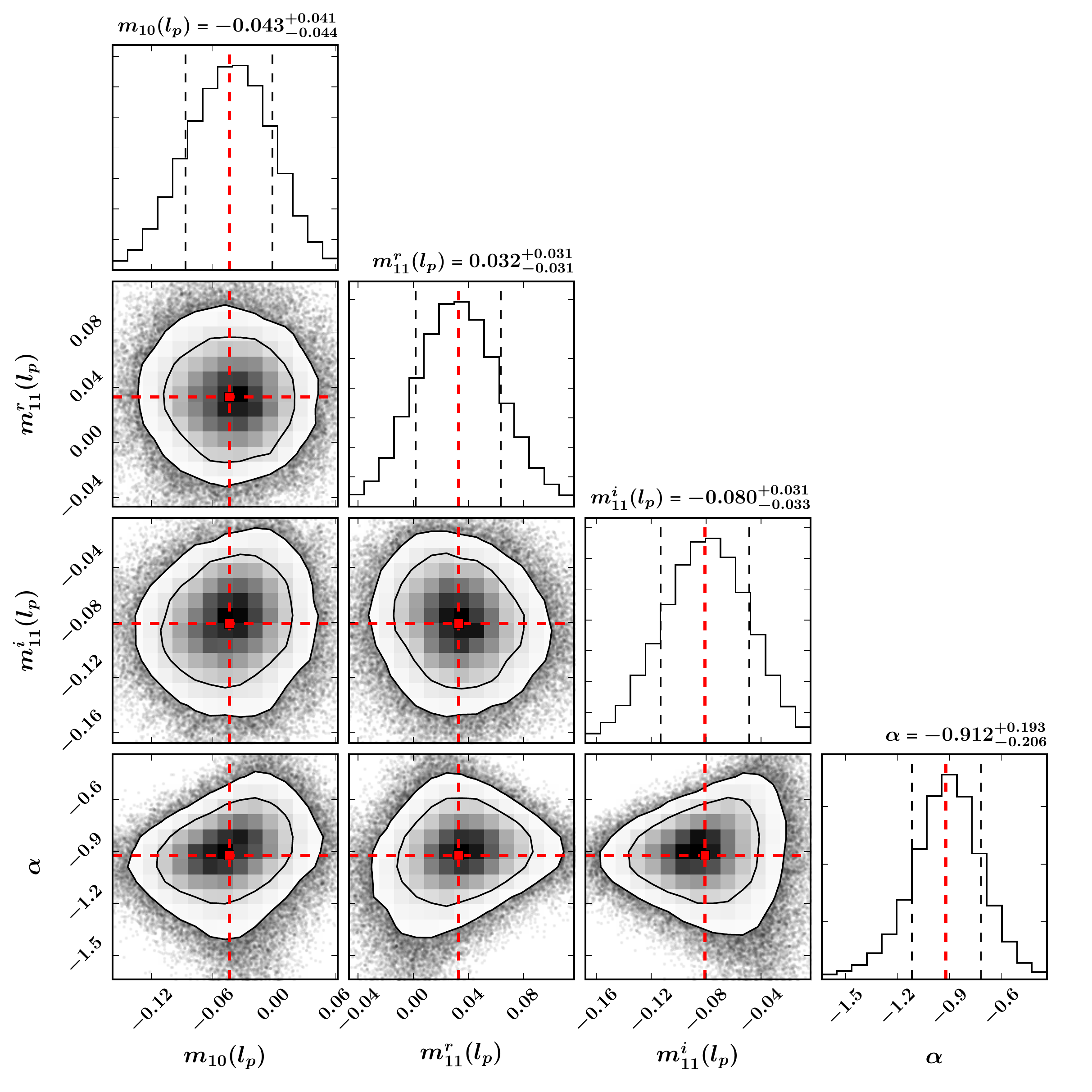}
\caption{The plot shows the joint and marginalized posterior distributions of parameters $m_{10}(l_p = 16)$, 
$m^{r}_{11}(l_p = 16)$, $m^{i}_{11}(l_p = 16)$ and $\alpha$ for the Planck SMICA map. Red dashed line 
indicates the mean of the sampled distribution. In 2D distributions, contours show 
regions of distribution containing 68\% and 90\% samples. In 1D distributions, black dashed lines mark 
16 and 84 percentiles of the distribution. The title above each histogram shows the median value 
and the 16 and 84 percentiles for the parameter. Respective mean and standard deviation of the 
parameters ($m_{10}(l_p), m^{r}_{11}(l_p), m^{i}_{11}(l_p), \alpha$) are $(-0.044, 0.043)$, $(0.033, 
0.031)$, $(-0.081, 0.032)$, $(-0.92, 0.22)$.}
\label{fig_smica_param_triangle}
\end{figure}

\begin{figure}[H]
\centering
\subfigure[ ]{\label{fig_smica_A} 
\includegraphics[width=0.8\textwidth]{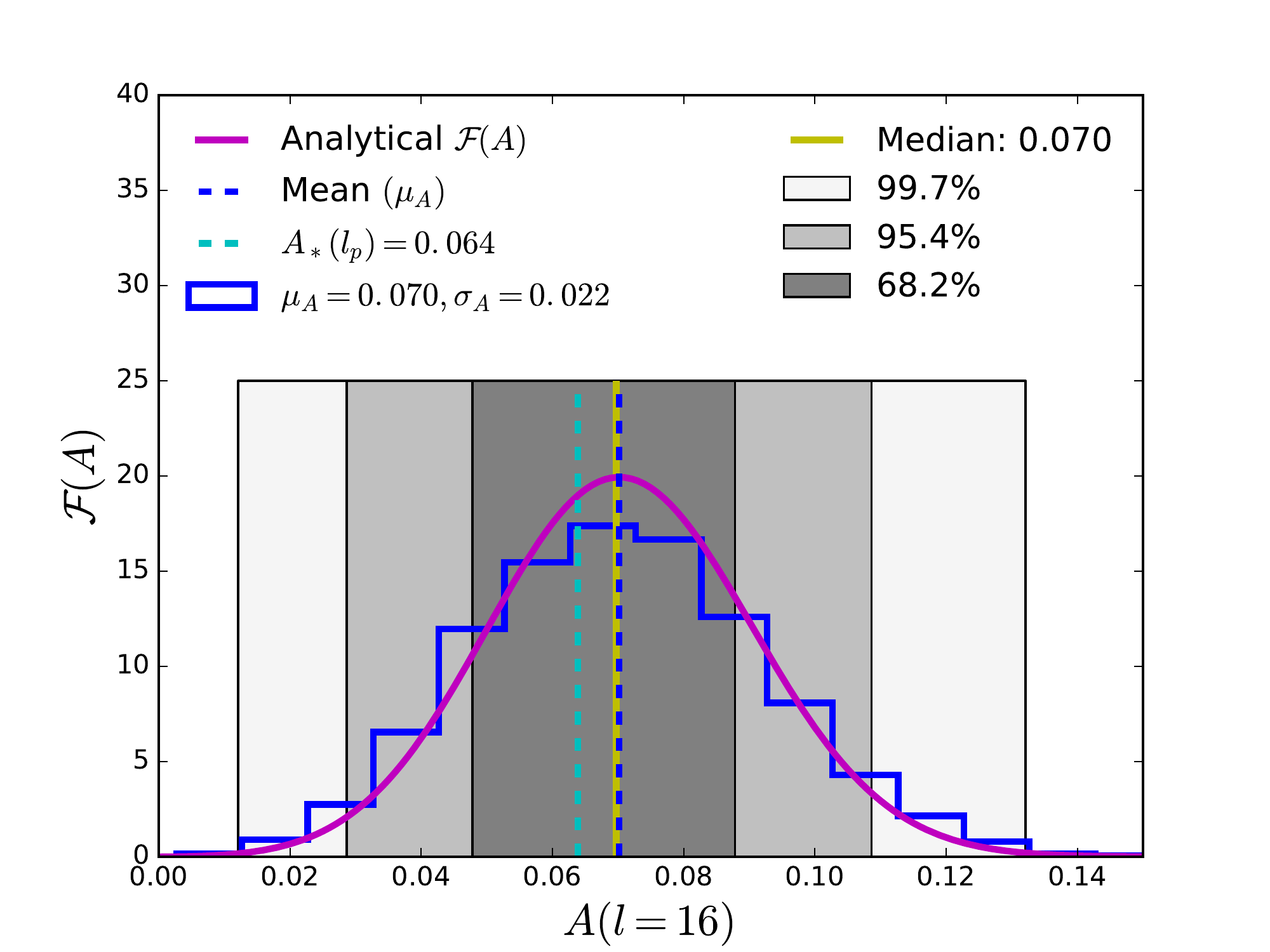}}
\subfigure[ ]{\label{fig_smica_theta_phi}
\includegraphics[width=0.8\textwidth]{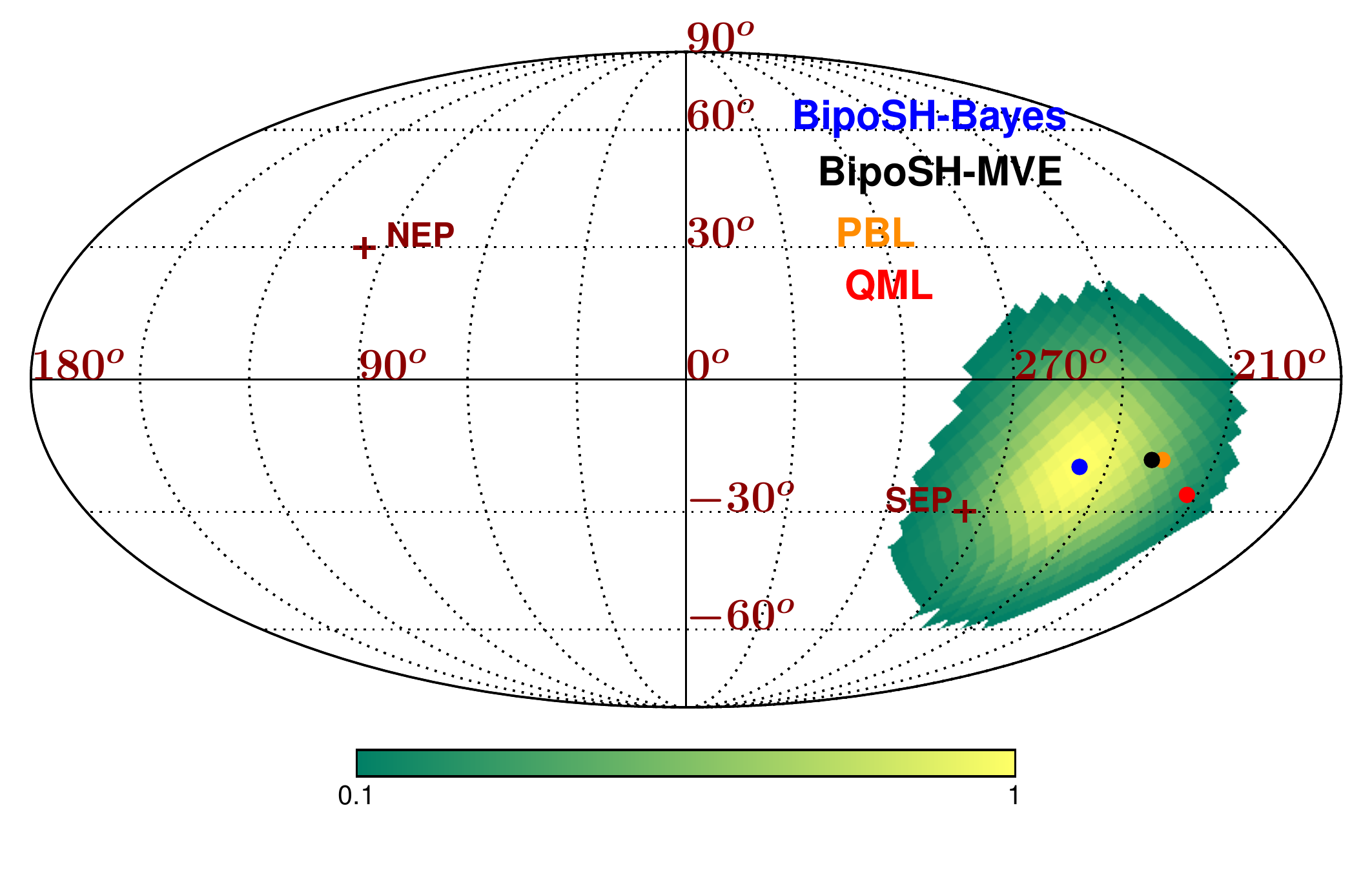}}
\caption{\textbf{(a)} Plot shows the distribution the dipole amplitude at pivot multipole $l = 16$. Dashed blue vertical line marks the mean $(\mu_{A})$ of the distribution, $\sigma_{A}$ is the standard deviation of the distribution. Cyan dashed line marks $A_*$, the dipole amplitude corresponding to the mean values of $m_{10}(l_p = 16)$, $m^r_{11}(l_p = 16)$, $m^i_{11}(l_p = 16)$. The magenta curve shows the expected analytical distribution of $A$, based solely on the mean and standard deviations of $m_{10}$, $m^{r}_{11}$, and $m^{i}_{11}$ given in figure \ref{fig_smica_param_triangle}. Functional form of this curve is given by $\calf(A)$ of \Eq{Prob_A_DM} with $A_* = 0.064$. \textbf{(b)} Figure shows the posterior distribution of $\theta_p$ and $\phi_p$ for SMICA map. $(\theta_p, \phi_p)$ values are 
binned using \texttt{HEALPix} \texttt{NSIDE} = 16 grid.
This histogram is normalized with respect to its peak value and is further smoothed by a Gaussian with standard deviation 3.7 degrees for presentation purpose. Blue dot represents the maximum a posteriori dipole direction (which corresponds to the direction obtained from the mean values of $m_{10}(l_p), m^r_{11}(l_p)$, and $m^i_{11}(l_p)$) called BipoSH-Bayes and has galactic coordinates $(l, b) = (247.8^o, -19.6^o)$. Also shown are the estimates from \cite{Planck_2015_isotropy}: (1) Pixel Based Likelihood (PBL) estimate (orange) $(l, b) = (225^o, -18^o)$, (2) Quadratic Maximum Likelihood (QML) estimate (red) $(l, b) = (213^o, -26^o)$, and (3) BipoSH Minimum Variance estimate (BipoSH-MVE) (black) $(l, b) = (228^o, -18^o)$.}
\end{figure}

The power-law model of dipole modulation is given by \Eq{eq_power_law_Al}. We perform the HMC analysis for the power-law model by taking the pivot point as $l_p= 16$ for NSIDE $=256$ SMICA temperature map. In figure \ref{fig_smica_param_triangle}, we show the posterior of $m_{10}(l_p = 16)$, $m^{r}_{11}(l_p = 16)$ and $m^{i}_{11}(l_p = 16)$ obtained from $10^{5}$ Monte Carlo samples after the Burn-In period of $5 \times 10^{4}$ samples. Figure \ref{fig_smica_param_triangle} also shows the posterior of the power law index $\alpha$ for the SMICA map. Similar to the analysis with the step model, the $m_{10}(l_p), m^r_{11}(l_p)$ and $m^i_{11}(l_p)$ parameters are treated as independent variables and correlation among these three parameters is not expected. This is evident from the joint distributions of these parameters given in the figure \ref{fig_smica_param_triangle}. However, the choice of pivot multipole $l_p$ can lead to  correlation between power law index $\alpha$ and ($m_{10}(l_p), m^r_{11}(l_p)$, $m^i_{11}(l_p)$). But our choice of pivot multipole $l_p = 16$ does not lead to any such significant correlation. 
The distribution of $A(l_p = 16)$ given in the figure \ref{fig_smica_A} reflects the combined effect of harmonic space parameters being away from the origin. The consistency of our result for a different choice of pivot point ($l_p =32$) is shown in Appendix \ref{alt_method}. 
The posterior distributions of \{$A(l_p = 16) , \theta_p, \phi_p$\} are obtained by transforming the Monte Carlo chains of \{$m_{10}(l_p = 16)$, $m^{r}_{11}(l_p = 16), m^{i}_{11}(l_p = 16)$\} to \{$A(l_p = 16) , \theta_p, \phi_p$\} using coordinate transformation formulas given in \Eq{m1M_to_Athephi}. Significant nonzero value of amplitude $A(l_p = 16)$ along with the nonzero value of $\alpha$ points to the presence of the dipole modulation with scale dependence.  The recovered power-law profile of the CHA signal is depicted in figure \ref{fig_smica_Al} along with the profile for the step-model and the Doppler Boost signal which is of the order $10^{-3}$ \cite{Riess:1995cg, 2003PhRvD..67f3001K,Planck_2013_DB,Challinor:2002zh, Amendola:2010ty, 2014PhRvD..89h3005M}. The recovered power-law profile indicates more than $0.5\%$ modulation effect up to the multipole of $l=256$ and hence is a contamination to the measurement of Doppler Boost signal at these low multipoles \cite{Planck_2013_DB}. The recovered profile can also lead to direction dependence in the cosmological parameters \cite{Mukherjee_dir_dep_param, Mukherjee_Wandelt_param}. Any theoretical studies in future to understand the origin of CHA need to satisfy the recovered profile shown in this analysis.

\begin{figure}[H]
\includegraphics[width=.8\linewidth, 
center]{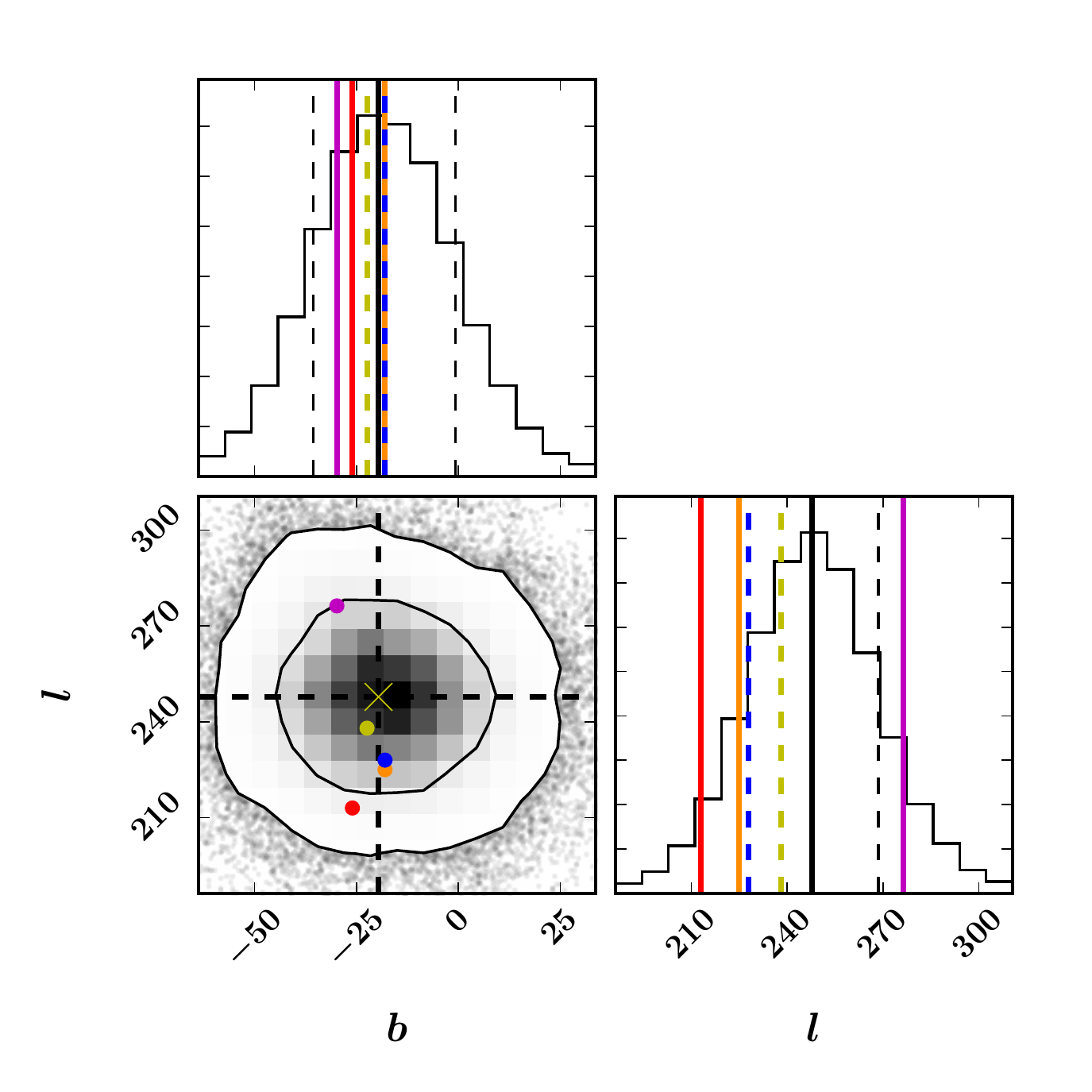}
\caption{The figure shows the joint and marginalized distributions of dipole direction $(\theta_p$ and $\phi_p)$, in Galactic coordinates, for power law model. In the joint distribution, yellow cross marks the peak of the distribution and contours mark regions containing 65\% and 95\% samples. For a quantitative comparison, we also plot our estimate of the dipole direction using step model (yellow dot) and the estimates from \cite{Planck_2015_isotropy}: (1) PBL estimate (orange) $(l, b) = (225^o, -18^o)$, (2) QML estimate (red) $(l, b) = (213^o, -26^o)$, and (3)BipoSH-MVE ( {blue}) $(l, b) = (228^o, -18^o)$. All these estimates are consistent within the 68\% confidence level except QML estimate which lies just outside the 68\% contour. The South Ecliptic Pole (SEP) is marked using magenta point and lies within the 68\% confidence region.}
\label{fig_smica_theta_phi_triangle}
\end{figure}

The distribution of the dipole direction for the SMICA map is given in the figure \ref{fig_smica_theta_phi}. The peak of the distribution gives our best-fit estimate of the dipole 
direction, which is $(l, b) = (247.8^o, -19.6^o)$ which is mildly deviant from the direction as reported by the previous analysis \cite{Planck_2013_isotropy,Planck_2015_isotropy}. We find that the direction of the dipole modulation under step model and power law model are in agreement and hence does not depend on the profile of the dipole modulation. The direction estimated under step model (see figure \ref{fig_smica_theta_phi_step_model}) 
$l, b = (238.0^o, -22.4^o)$ is consistent with that estimated under power law model 
$l, b = (247.8^o, -19.6^o)$ (see figure \ref{fig_smica_theta_phi}).
A more quantitative comparison between the direction recovered in our analysis and the direction reported in the literature is given in figure \ref{fig_smica_theta_phi_triangle}. The estimates of the dipole direction provided by Pixel Based Likelihood (PBL) estimate and BipoSH minimum variance estimate (BipoSH-MVE) are consistent within the 68\% confidence level, whereas Quadratic Maximum Likelihood (QML) estimate of the dipole direction lies outside the 68\% contour. In particular, we note that the South Ecliptic Pole (SEP) falls within the boundary of the 68\% confidence region.

\begin{figure}[h]
 \includegraphics[width=0.8\linewidth,center]{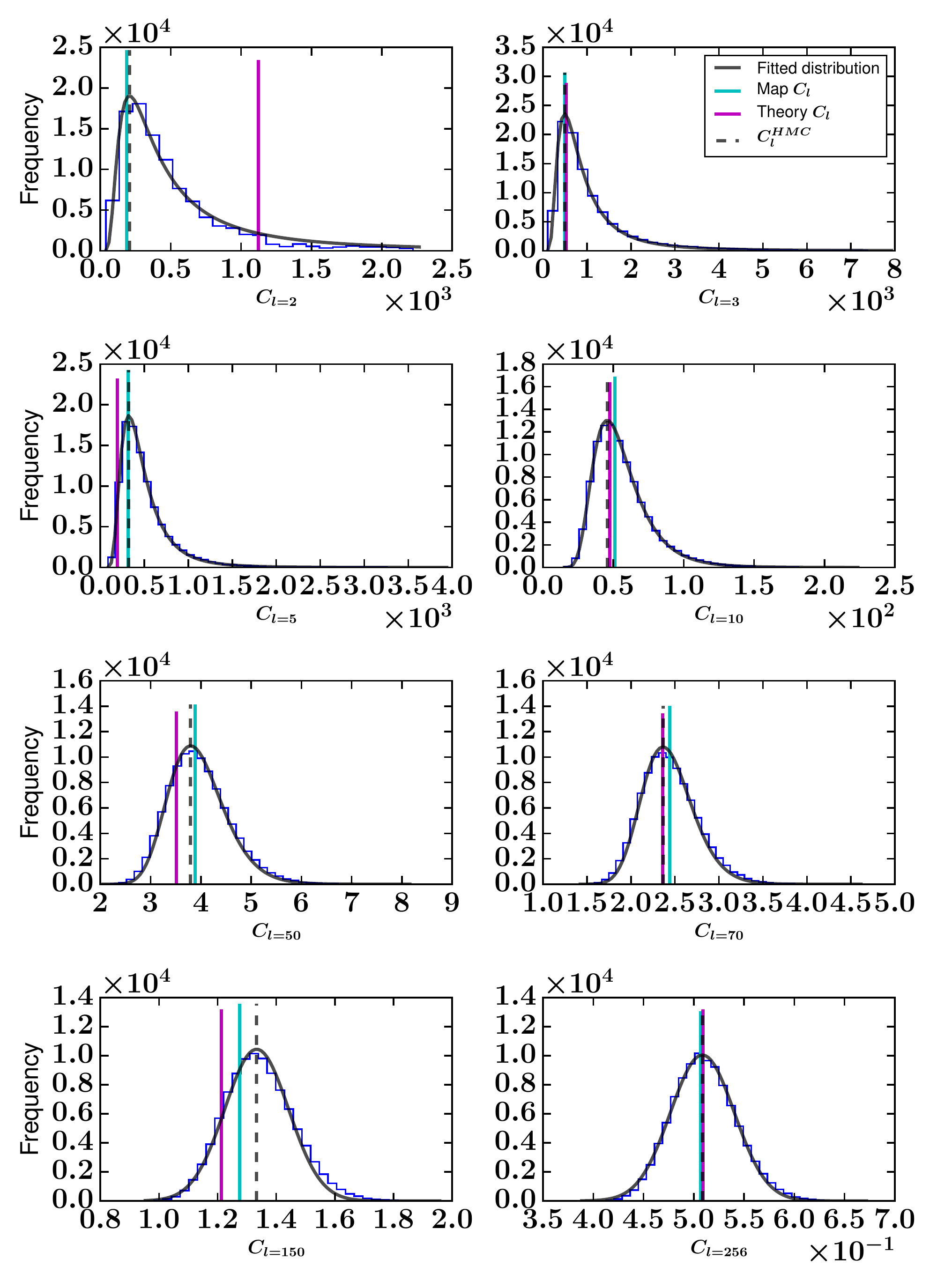}
\caption{Figure shows the posterior distributions for the angular power spectra of SMICA map at select multipoles. In blue, we show the histogram of the Monte Carlo samples of $C_l$. Black curve shows the fitted distribution to these histograms. Black dashed line indicates the peak of the fitted distribution, 
$C^{HMC}_l$. Cyan 
line marks the $C_l$ of the map realization. Magenta line shows the $\Lambda$CDM best-fit theory angular power 
spectrum provided by the \textit{Planck} collaboration.}
\label{fig_smica_Cl_subplots}
\end{figure}

\begin{figure}
\includegraphics[width=\linewidth,center]
{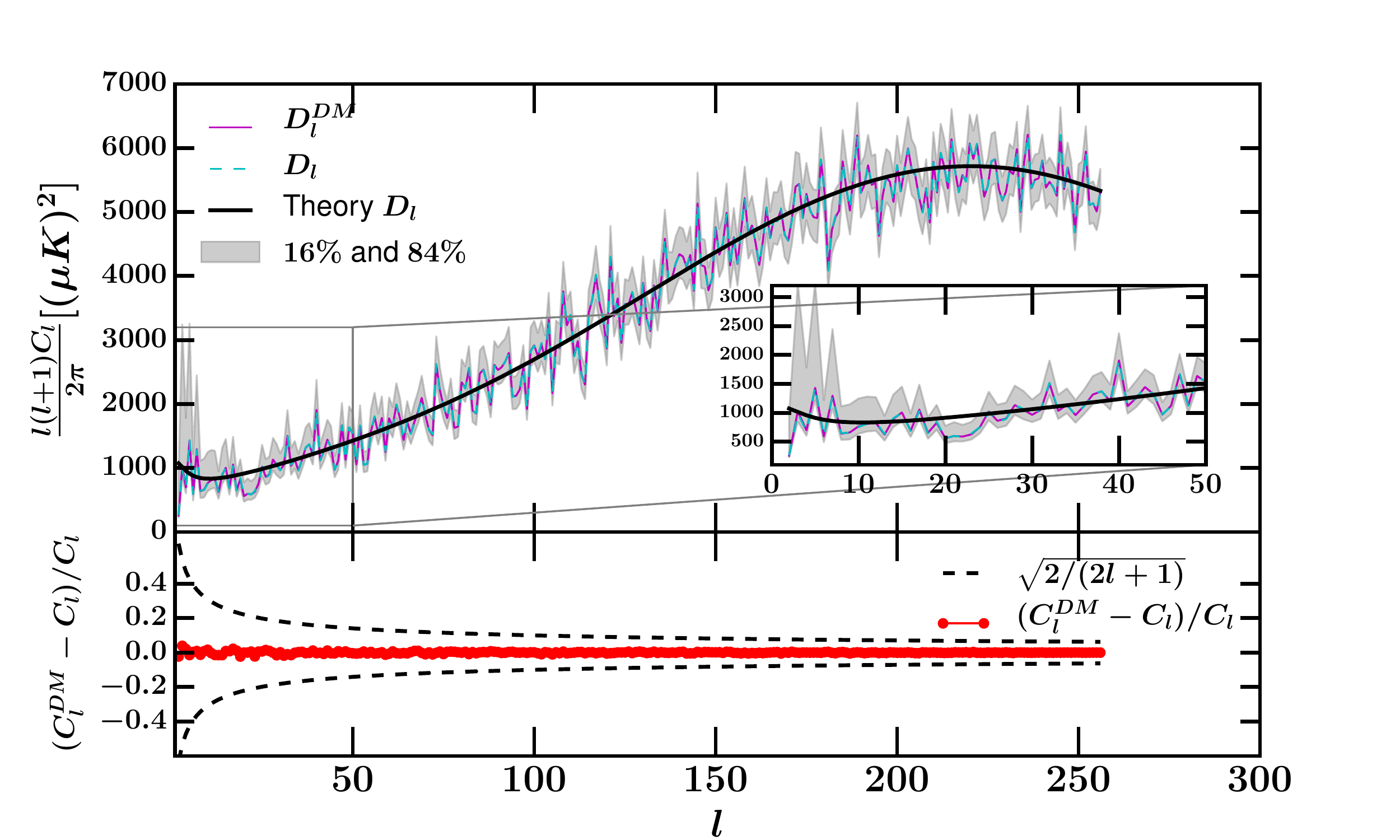}
\caption{The upper panel of the figure shows the maximum a posteriori estimate of the 
angular power spectrum of joint analysis with dipole modulation parameters ($C^{DM}_l$) and that of the analysis 
without dipole modulation parameters ($C_{l}$), for the SMICA map. Black curve shows the $\Lambda$CDM best-fit theory angular power spectrum provided by the \textit{Planck}. The gray shaded region marks the region of 16 percentiles and 84 percentiles of the $C^{DM}_l$ distribution. In bottom panel, we plot the relative difference between $C^{DM}_l$ and $C_{l}$ and compare it with the quantity $\sqrt{2/(2l + 1)}$,  {shown by black dashed line}. $D_l$ stands for $l(l+1)C_l/(2\pi)$.}
\label{fig_smica_Cl}
\end{figure}

In remaining part of this section, we discuss the probability distribution of $C_l$ obtained using HMC analysis of the SMICA map. 
In figure \ref{fig_smica_Cl_subplots} we show the posterior distribution for $C_l$ jointly estimated 
with the dipole modulation parameters for SMICA map at select multipoles. We fit the analytical probability distribution of $C_l$ given 
in \cite{Percival:2006ss} to the histogram of $C_l$ samples. We use \texttt{SciPy} 
routine \texttt{scipy.optimise.curve\_fit} to implement the fitting \cite{scipy_tools}. After fitting the 
probability distribution of $C_l$ to the histogram of the samples of $C_l$, we get the best-fit 
estimate of the angular power spectrum ($C^{HMC}_l$), which matches very well with the quadratic estimation of the angular power spectrum obtained from the map. In figure \ref{fig_smica_Cl} we show two estimates of the angular power spectrum: (1) angular power spectrum estimated jointly with the dipole modulation parameters ($C^{DM}_l$) and (2) angular power spectrum estimated without the dipole modulation parameters ($C_l$). Both the estimates are compared with the best-fit $\Lambda$CDM angular power spectrum provided by \textit{Planck} \cite{Planck_2015_Cl}. We note that irrespective of whether we estimate $C_l$ with or without dipole modulation parameters, the estimates of $C_l$ do not show any significant differences. The suppression in the power spectrum in the multipole range $l\in\{20,30\}$ remains at the same statistical significance even in the joint analysis of $A(l)$ and $C_l$. The estimated CHA signal also remains unaltered when the $C_l$ are kept fixed at the fiducial $\Lambda$CDM values. The corresponding results are shown in Appendix \ref{Appendix_step_model_SMICA}.  

\begin{figure}
\centering
\includegraphics[width=0.5\linewidth]{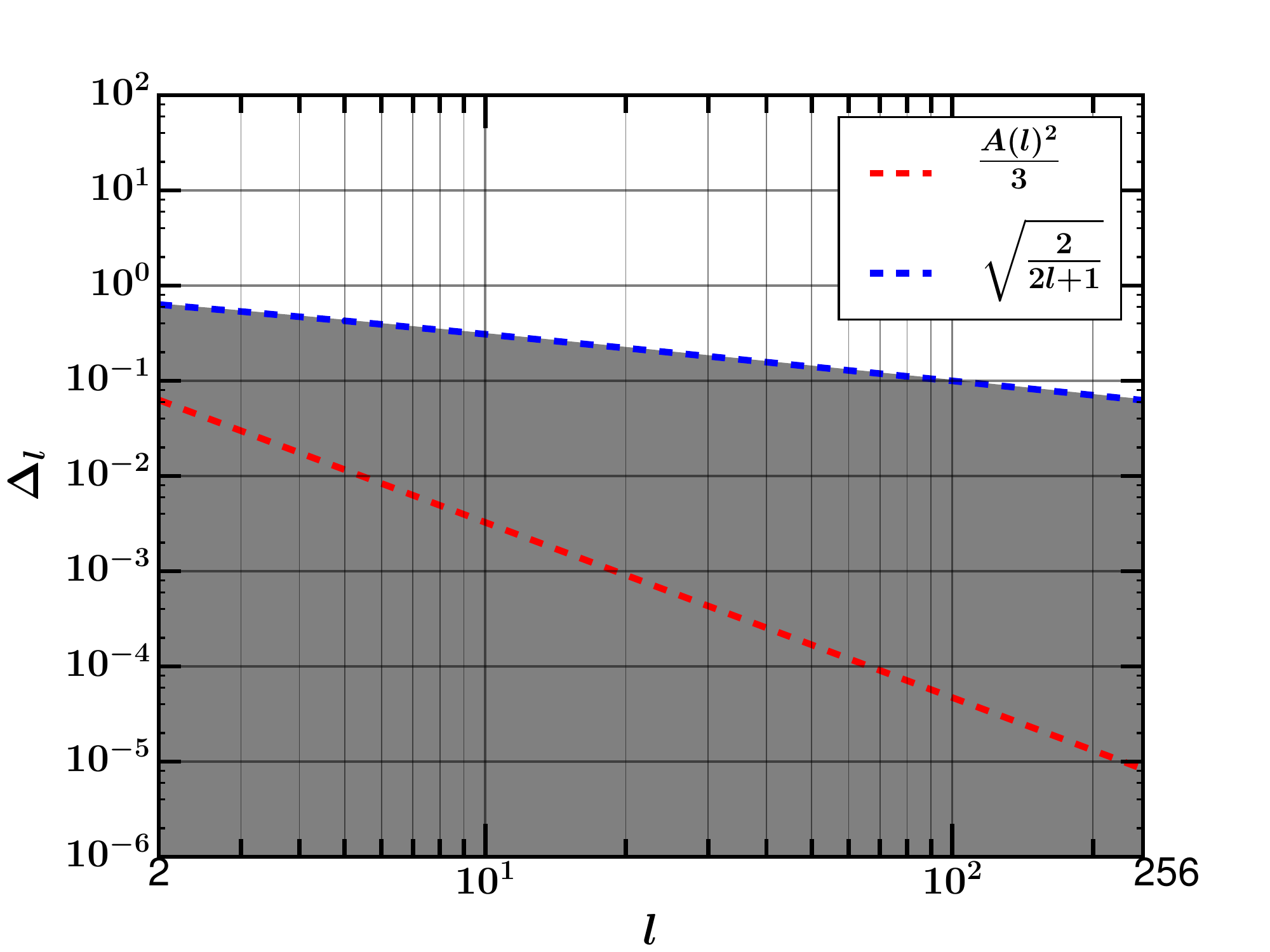}
\caption{The relative correction to the angular power spectrum $(\Delta_l)$ due to the second order effect of dipole modulation amplitude (see \Eq{Delta_Cl}) compared to the Cosmic Variance of $C_l$.  {Red dashed line} depicts the relative contribution to $C_l$. Blue  {dashed line} curve depicts the level $\sqrt(2/(2l+1))$.}\label{fig_smica_Al_tildeCl}
\end{figure}

The recovered power-law of the dipole modulation signal depicted in figure \ref{fig_smica_Al} shows a strong modulation field in the low $l$. As a result, the amplitude of the second order terms $A^2(l)$ gets stronger at the low $l$ (as mentioned in \Eq{cov_mat}). 
We estimate the correction to the angular power spectrum due to second order term and depict the relative correction to the angular power spectrum $\Delta_l$ in figure \ref{fig_smica_Al_tildeCl}. For comparison, we also show the $1\sigma$ cosmic variance error-bar by the shaded region. This establishes that the second order term is negligible and does not play any significant role in the analysis.

\section{Model comparison between SI and non-SI models of CMB map: Estimation of Bayes factor}\label{model_comp}
\paragraph{Bayesian model comparison:}
Bayesian analysis allows comparison of two models (say $M_0$ and $M_1$). One compares the
probability of the model $M_0$ given data ($\calP(M_0|d)$) and the probability of the model $M_1$ 
given data ($\calP(M_1|d)$) using the following ratio called Bayes factor 
\begin{equation}\label{def_bayes_factor}
 \frac{\calP(M_0|d)}{\calP(M_1|d)} = \frac{\mathcal{E}(d|M_0) \calP(M_0)}{\mathcal{E}(d|M_1) \calP(M_1)}.
\end{equation}
In the above equation, $\mathcal{E}(d|M)$ stands for the probability of the data given model $M$ and $\mathcal{E}(d|M)$ is 
called Bayesian Evidence for $M$. $\calP(M)$ stands for the prior probability of the model $M$.
In the situation where two models are considered a priori equally probable ($\calP(M_0) = \calP(M_1)$), the Bayes 
factor is the ratio of the evidence of the models under the same data. If the Bayes factor 
is close to unity, the inference is that the data do not discriminate between the two models.

In our work, we compare the SI model ($M_{SI}$) of CMB with the power-law dipole modulation 
model ($M_{DM}$). The Bayes factor in favor of the SI model, $B_{SI-DM}$ is 
\begin{equation}\label{B_01}
 B_{SI-DM} \equiv \frac{\mathcal{E}(d| M_{SI}) }{\mathcal{E}(d | M_{DM}) },
\end{equation}
where we have assumed that prior probability $\calP(M_{SI}) = \calP(M_{DM})$.
In above equation, $\mathcal{E}(d| M_{SI})$ and $\mathcal{E}(d| M_{DM})$ are evidence under  $M_{SI}$ and $M_{DM}$, respectively, which is defined as
\begin{equation}\label{evidence_nSI}
\mathcal{E}(d|M) = \int \calL(w_x, w_y, w_z, \alpha ) \Pi(w_x, w_y, w_z, \alpha | M) dw_xdw_y dw_z d\alpha,
\end{equation}
where, $w_x$, $w_y$, and $w_z$ are convenient re-parameterizations of $m^r_{11}$, $m^i_{11}$, and $m_{10}$ respectively, as defined 
in \Eq{def_wxyz}. $\calL(w_x, w_y, w_z, \alpha )$ denotes the likelihood and $\Pi(w_x, w_y, w_z, \alpha | M)$ is the prior on parameters.
\paragraph{Bayes factor using Savage-Dickey density ratio:}
The power-law dipole modulation model has four additional parameters compared to the SI model, which are $w_x(l_p), w_y(l_p), w_z(l_p)$, and $\alpha$. The model $M_{SI}$ is nested within the model $M_{DM}$ in the sense that using $(w_x(l_p), w_y(l_p), w_z(l_p)) = (0,0,0)$ reduces $M_{DM}$ to $M_{SI}$ for all values of $\alpha$. The Bayes factor for nested models can be obtained by using Savage-Dickey density ratio (SDDR) \cite{dickey1971_2, Trotta_BayesSky}. For SI model nested in the power law dipole modulation model, the SDDR is
\begin{equation}\label{SD_ratio_1}
B_{SI-DM} = \frac{\calP_{C_l}(w_x = 0, w_y = 0, w_z = 0,\alpha |d,DM)}{\Pi(w_x = 0, w_y = 0, w_z = 0, \alpha | DM)},
\end{equation}
where $\calP_{C_l}(w_x = 0, w_y = 0, w_z = 0,\alpha |d,DM)$ is posterior marginalized over $C_l$. $\alpha$ dependence of SDDR in \Eq{SD_ratio_1} is weak because the likelihood conditioned at $(w_x, w_y, w_z) = (0,0,0)$ is independent of $\alpha$. The prior mentioned in \eqref{SD_ratio_1} can be written as \begin{equation}\label{eq_prior_condtional}
\Pi(w_x, w_y, w_z, \alpha | M_{DM}) = \Pi(w_x, w_y, w_z |\alpha; M_{DM}) \Pi(\alpha | M_{DM}).
\end{equation}
We choose uniform prior on $\alpha$  over the range $\alpha = -2 \text{ to } 0$. Hence, the normalized prior density of $\alpha$ is
\begin{equation}\label{prior_alpha}
 \Pi(\alpha | M_{DM}) = \frac{1}{2}\,\, \text{for $-2 \leq \alpha \leq 0$}.
\end{equation}
The amplitude of the dipole modulation $A(l)$, decides the relative magnitude of the off-diagonal terms in the covariance matrix compared to the diagonal terms ($C_l$). In order to ensure the positive definiteness of the covariance matrix
\begin{equation}\label{prior_Al}
 A(l) < \frac{1}{2} \quad \quad \forall \quad l.
\end{equation}
In spherical polar coordinates, with $r$ as defined in \Eq{def_r}, the above argument gives an upper bound: $\mathbf{r \leq  
\sqrt{(\pi/3)} \approx 1}$, which is denoted by $R$. To obey \Eq{prior_Al}, we need to make the upper bound on $r$ dependent on $\alpha$. 
Otherwise, for low values of $\alpha$ close to $-2$, $A(l)$ can become greater than $1/2$ at low multipoles for sufficiently high $A(l_p)$. Hence, $\alpha$ dependent upper bound $R(\alpha)$ is
\begin{equation}\label{R_alpha}
 R(\alpha) = \frac{1}{2} \Big( \frac{l}{l_p} \Big)^{-\alpha} \sqrt{\frac{4 \pi}{3}} = 
 1.02 \Big( \frac{l}{l_p} \Big)^{-\alpha}.
\end{equation}
At $\alpha = -2$, with $l_p = 16$ and $l = 2$, $R(\alpha) = 0.016$.   

We choose a prior on $(w_x, w_y, w_z)$ which has uniform density within the sphere of radius $R(\alpha)$, centred at $(w_x, w_y, w_z) = (0,0,0)$ and zero outside. Normalized form of such prior is
\begin{equation}\label{Prior_xyz}
\Pi_{1}(w_x, w_y, w_z|\alpha, DM) = 
\begin{cases}
\frac{3}{4 \pi R^3(\alpha)} & \text{if } \sqrt{w_x^2 + w_y^2 + w_z^2} \leq R(\alpha) \\
    0              & \text{otherwise}.
\end{cases}
\end{equation}
Note that the prior density is independent of the direction. The form of prior in \Eq{Prior_xyz} ensures positive definiteness of the covariance matrix at all scales for the range of $\alpha$ considered in the analysis.
Choosing an uniform density prior as given in \Eq{Prior_xyz} implies that the angle marginalized prior density of the amplitude is
\begin{equation}
\Pi_1(r | \alpha, DM) = \frac{3r^2}{R^3(\alpha)} \quad \text{for} \quad r \leq R(\alpha).
\end{equation}
As mentioned in section \ref{3p1}, we have sampled the likelihood distribution without augmenting it with any prior. We obtain the posterior by weighting the likelihood samples by the prior density given in \Eq{Prior_xyz}. With this procedure, the SDDR in favour of the SI model is found to be
\begin{equation}
    B_{SI-DM} \approx 0.4.
\end{equation}
In Appendix \ref{Prior_choices}, we obtain the SDDR for two more choices of prior densities and show that the variations are negligible.

This implies that the data do not favor the concordance model of cosmology over the non-SI model. The large prior range for the non-SI parameters \{${A(l), \alpha}$\} is the key reason for the inconclusive Bayes factor, though the likelihood clearly peaks away from zero (the expected value under the concordance model of cosmology) as shown in figure \ref{fig_smica_A}. In summary, \textit{the best available nearly full sky CMB temperature data does not rule out non-SI cosmological model even after carefully including the observational non-idealities (masking, anisotropic noise, instrument beam). Other cosmological probes like CMB polarization, weak lensing, and large scale structure surveys keep the hope alive to provide conclusive evidence for or against this enigmatic anomaly in the future.}

\section{Discussions and Conclusions}\label{Conclusion}
In this paper, we study the combination of two  large angular scale CMB anomalies, namely the power suppression and dipole power asymmetry (also called the Cosmic Hemispherical Asymmetry (CHA)) in the Bayesian framework using the Planck-2015 SMICA temperature map. Most of the analysis in the literature assume a step model which have a non-zero  
amplitude of the dipole signal only below an arbitrarily chosen value of multipole ($l_{cut}$). This value of the $l_{cut}$ is usually chosen as $64$, though a few analysis are also performed setting $l_{cut}=600$ \cite{Flender:2013jja}.
We model the amplitude of the modulation dipole
with a two-parameter power-law, the amplitude at the pivot multipole ($A(l_p)$) and the spectral index ($\alpha$). The parameters $\theta_p$ and $\phi_p$ give the direction of the modulation dipole.
The joint estimate of these four parameters in harmonic space together with the temperature coefficients of the map $a_{lm}$ and power spectrum $C_l$ are performed after considering the observational non-idealities like the partial sky, instrument beam, and anisotropic instrument noise.

The jointly estimated angular power spectrum $C_l$ matches well with the value of $C_l$ reported by Planck \cite{Planck_2015_Cl}. The low value of the quadrupole, suppression of power between the multipole range $20-30$ holds at the same significance. Hence, we reconfirm the existence of power suppression from this joint study even within the more general statistical model including an anisotropic covariance matrix.

Our estimate reveals a scale-dependent CHA signal with the maximum a posteriori values of the parameters as $A_*(l_p = 16) = 0.064 \pm 0.022$ and $\alpha = -0.92 \pm 0.22$ for a power law parametrization. The direction of the maximum signal in the galactic coordinates is $(l,b) = (247.8^o, -19.6^o)$. 
The spectral index of the power-law differs by about one $\sigma$ from the previously reported value by Aiola et al. \cite{Aiola_2015_power_law}, when compared with their result for  $l_{max}=300$. However, the amplitude at the pivot multipole $l_p = 16$ ($A(l_p = 16) = 0.071$) is consistent within one sigma of our result. These mild differences can arise due to the choice of different pivot points, the maximum value of the multipoles used ($l_{max}$), use of the different Planck noise simulations  (FFP6 instead of FFP8) and most importantly due to the difference in the analysis framework. The direction of the asymmetry is also recovered consistently in both the analysis. We note that, the best fit direction of the dipole is almost the same for both step-model and power-law model analysis. This is expected because both models differ only in the treatment of the dipole amplitude. 

We also make a model comparison between power law form of the scale-dependent dipole modulation model and the concordance SI model of CMB. The estimated Bayes factor in favor of the concordance model is $0.4$ and does not rule strongly in favor of SI model, hence remains inconclusive. The current best available temperature data does not stand out loud in favor of the concordance model and shows only a mild inclination in favor of the non-SI model. Future datasets from CMB-polarization, weak lensing and large scale structure are capable to shed light to this enigmatic anomaly.

Though the two-parameter power-law model is  simplistic, it has the ability to capture the scaling of scale dependence. A future study with the additional parameters beyond the power-law form can be performed with the availability of the CMB polarization and large scale structure datasets.
The extrapolation of the retrieved power-law form to small angular scales leads to a CHA signal less than $0.1\%$ for angular scales below $10$-arcmin. This is sub-dominant (as required) from the well-predicted Doppler Boost signal \cite{Challinor:2002zh, Amendola:2010ty, 2014PhRvD..89h3005M} due to our local motion with a velocity $\beta\equiv v/c= 1.23 \times 10^{-3}$ \cite{Riess:1995cg, 2003PhRvD..67f3001K}, which is also measured by Planck using the high resolution temperature data \cite{Planck_2013_DB}. 
Hence, the retrieved signal profile obtained in this paper satisfies the observational requirement at both small and large angular scales. The profile of the power-law signal obtained in this analysis will be useful guide in any pursuit of a theoretical understanding of CHA signal and also to make predictions for other cosmological probes including CMB polarization \cite{Dvorkin:2007jp,Zibin:2015ccn, Contreras:2017zjv, Mukherjee:2015wra, Ghosh:2018apx, Ghosh:2016tbj}. Bayesian inference of the BipoSH coefficients using HMC discussed in this work can be generalized to CMB polarization. The systematic noise due to the instrument and foreground effects become more important for low multipole polarization data provided by \textit{Planck} and including these effects accurately requires an in-depth analysis that we defer to future work.

\vskip20pt
\textbf{Acknowledgements:}
S.S. acknowledges University Grants Commission (UGC), India for providing the financial support  as 
Senior Research Fellow. S. M. and B. D. W. acknowledge the support of Simons Foundation for this work. This work is also supported by the Labex ILP (reference ANR-10-LABX-63) part of the Idex SUPER, and received financial state aid managed by the Agence Nationale de la Recherche, as part of the programme Investissements d'avenir under the reference ANR-11-IDEX-0004-02. Authors would like to thank Planck collaboration and NERSC for providing the FFP 
simulations and other auxiliary files. The present work is carried out using the High Performance 
Computing facility at IUCAA \footnote{\url{http://hpc.iucaa.in}}. Some of the results in this paper have been derived using the HEALPix package 
\cite{Gorski_healpix_paper}. We also acknowledge the use of \texttt{CAMB}~\cite{Lewis:1999bs}, 
\texttt{IPython}~\cite{PER-GRA:2007}, Python packages \texttt{Matplotlib}~\cite{Hunter:2007}, 
\texttt{NumPy}~\cite{numpy}, \texttt{SciPy}\cite{scipy_tools} and \texttt{corner}~\cite{corner}.

\appendix
\section{Derivation of \texorpdfstring{$\calP(C_l,m_{1M}|\bd)$}{Lg}}
\label{appendix_P_Cl_m1M}
This appendix provides a detailed derivation of \Eq{P_of_S_given_d} and \Eq{joint_P_cl_m10}. To derive 
\Eq{P_of_S_given_d}, we marginalize over $\ba$ from \Eq{Prob_a_S}. We simplify the following expression in \Eq{Prob_a_S}
\begin{equation}
E = (\textbf{d} - \textbf{a})^{\dagger} \textbf{N}^{-1}  (\textbf{d} -\textbf{a}) 
+ \textbf{a}^{\dagger} \textbf{S}^{-1} \textbf{a},
\end{equation}
by completing the square \cite{1992ApJ...398..169R}.
\begin{align*}
E = \bd^{\dagger}(\bS + \bN)^{-1}\bd + 
(\ba - \tilde \ba)^{\dagger}(\bS^{-1} + \bN^{-1}) (\ba - \tilde \ba),
\end{align*}
where $\tilde \ba = \bS(\bS + \bN)^{-1}\bd$ is Wiener filter applied to the data.
\Eq{Prob_a_S} now becomes
\begin{equation}\label{Prob_a_S_revised}
 \calP(\bS, \ba | \bd) = \frac{1}{\sqrt{|\bN| |\bS| } (2 \pi)^n} 
 \exp {\Big[-\frac{1}{2}(\ba - \tilde \ba)^{\dagger}(\bS^{-1} + \bN^{-1})(\ba - \tilde \ba) \Big] }
 \exp { \Big[ -\frac{1}{2} \bd^{\dagger}(\bS + \bN)^{-1}\bd \Big ]}.
\end{equation}
Using the results
\begin{equation}
 \int \exp \Big[ -\frac{1}{2} (\ba - \tilde \ba)^{\dagger} (\bS^{-1} + \bN^{-1}) 
 (\ba - \tilde \ba) \Big] d\ba = (2 \pi)^{n/2} \sqrt{|(\bS^{-1} + \bN^{-1})^{-1}|},
\end{equation}
and 
\begin{equation}
 |(\bS^{-1} + \bN^{-1})| = \frac{|(\bS + \bN)|}{|\bS||\bN|},
\end{equation}
in \Eq{Prob_a_S} to get $P(\textbf{S}| \textbf{d})$ as given in \Eq{P_of_S_given_d}. Now we derive 
the joint probability distribution $C_l$ and $m_{10}$ given in \Eq{joint_P_cl_m10} from \Eq{P_of_S_given_d}. We write the sum of signal covariance matrix $\bS$ and noise covariance matrix $\bN$ in the following manner
\begin{equation}
 \bS + \bN = \bD + m_{10}\bO_1.
\end{equation}
Since we assume $\bN$ to be diagonal, it contributes only to $\bD$. $\bO_1$ is the off-diagonal 
part of $\bS$ without $m_{10}$. Taylor series expansion of inverse of the sum $\bS + \bN$, upto 
second order in $m_{10}$ gives
\begin{equation}\label{matrix_sum_inverse}
 (\bS + \bN)^{-1} = (\bD + m_{10}\bO_1)^{-1} = 
 \bD^{-1} - m_{10} \bD^{-1}\bO_1 \bD^{-1} + m^2_{10} (\bD^{-1}\bO_1)^2 \bD^{-1}.
\end{equation}
Determinant $|\bS + \bN|$ in \Eq{P_of_S_given_d} also has the $m_{10}$ dependence expressed as 
\begin{equation}
 |\bS + \bN| = |\bD| (|\bI + m_{10} \bD^{-1} \bO_1|).
\end{equation}
Making use of the matrix identity $\ln |M| = Tr[\ln (M)]$ to compute the logarithm of the 
determinant, we get
\begin{equation}
 \ln(|\bI + m_{10} \bD^{-1} \bO_1|) = Tr[\ln(\bI + m_{10} \bD^{-1} \bO_1)].
\end{equation}
Taylor series expansion of logarithm gives
\begin{equation}
 \ln(\bI + m_{10} \bD^{-1} \bO_1) = \sum^{\infty}_{k = 1} \frac{(-1)^{k+1}}{k} (m_{10} \bD^{-1}\bO_1)^k.
\end{equation}
Then, keeping the terms up to second order in $m_{10}$, we get 
\begin{equation}
 \ln(\bI + m_{10} \bD^{-1} \bO_1) = m_{10} \bD^{-1}\bO_1 -\frac{m^2_{10}}{2} (\bD^{-1}\bO_1)^2.
\end{equation}
\begin{equation}
 Tr[\ln(\bI + m_{10} \bD^{-1} \bO_1)]  = 
 m_{10} Tr[\bD^{-1}\bO_1] - \frac{m^2_{10}}{2} Tr[(\bD^{-1}\bO_1)^2].
\end{equation}
\begin{equation}
\ln(|\bI + m_{10} \bD^{-1} \bO_1|)  = 
- \frac{m^2_{10}}{2} Tr[(\bD^{-1}\bO_1)^2] \qquad\qquad{\because Tr[\bD^{-1}\bO_1] = 0}.
\end{equation}
\begin{equation}\label{matrix_log_det}
 \ln(|\bS + \bN|) = \ln(|\bD|) - \frac{m^2_{10}}{2} Tr[(\bD^{-1}\bO_1)^2].
\end{equation}
We substitute \Eq{matrix_sum_inverse} and \Eq{matrix_log_det} in the following form of 
\Eq{P_of_S_given_d}
\begin{equation}
 2 \ln(\calP(\bS|\bd)) = 2 \ln(\calP(C_l, m_{10} |\bd)) = 
 - n \ln(2\pi) - \ln|\bS + \bN| - \bd^{\dagger}(\bS + \bN)^{-1}\bd,
\end{equation}
to obtain
\begin{eqnarray}\label{P_Cl_m10_semifinal_eq}
 &-& 2 \ln(\calP(C_l, m_{10} |\bd)) = 
 n \ln(2\pi) + \ln|\bD| + \bd^{\dagger} \bD^{-1}\bd \nonumber\\
 &-& m_{10} \bd^{\dagger} \bD^{-1} \bO_1 \bD^{-1} \bd - 
 \frac{m^2_{10}}{2} Tr[(\bD^{-1}\bO_1)^2] 
 + m^2_{10} \bd^{\dagger} (\bD^{-1} \bO_1)^2 \bD^{-1} \bd.
\end{eqnarray}
\Eq{P_Cl_m10_semifinal_eq} can be further rearranged in following manner
\begin{eqnarray}\label{P_Cl_m10_final_eq}
 - \ln(\calP(C_l, m_{10} |\bd)) 
 &=& \frac{n}{2} \ln(2\pi) + \frac{1}{2}\ln|\bD| + \frac{1}{2}\bd^{\dagger} \bD^{-1}\bd \nonumber\\
 &+& \frac{Tr[(\bD^{-1}\bO_1)^2]}{4} M \Big\{ m^2_{10} - 2 m_{10} 
 \frac{\bd^{\dagger} \bD^{-1} \bO_1 \bD^{-1} \bd}{Tr[(\bD^{-1}\bO_1)^2]} M^{-1}\Big\},
\end{eqnarray}
where $M$ represents following quantity
\begin{equation}
 M = \Big[ \frac{2 \bd^{\dagger} (\bD^{-1} \bO_1)^2 \bD^{-1} \bd}{Tr[(\bD^{-1}\bO_1)^2]} - 1 \Big].
\end{equation}
 \Eq{joint_P_cl_m10} follows from \Eq{P_Cl_m10_final_eq} given above.

\section{Position and Momentum derivatives}
\label{Appendix_PALMll_dot}
This appendix contains the discussion of the position and momentum derivatives required for the Hamiltonian dynamics in HMC.

Following are the Hamilton's equations for $(a^r_{lm}, a^i_{lm})$
\begin{equation}
 \dot{a^r}_{lm} = \frac{p^r_{lm}}{\mu^r_{lm}}, \quad \dot{p^r}_{lm} = -\frac{\partial \calH}{\partial 
a^r_{lm}}, ~
 \dot{a^i}_{lm} = \frac{p^i_{lm}}{\mu^i_{lm}}, \quad \text{and} \quad \dot{p^i}_{lm} = -\frac{\partial \calH}{\partial 
a^i_{lm}}.
\end{equation}
This involves time derivative of conjugate momentum corresponding to $a_{lm}$
\begin{equation}\label{palm_dot}
 \dot{p}_{lm} = -\frac{1}{2} \frac{\pd \ba^{\dagger}\bS^{-1} \ba }{\pd a_{lm}} - 
 \frac{1}{2} \frac{\pd [(\bd - \ba)^{\dagger} \bN^{-1} (\bd - \ba)]}{\pd a_{lm}}.
\end{equation}
After taking derivative with respect to $a_{lm}$, above equation reduces to
\begin{equation}\label{palm_dot_2}
 \dot{p}_{lm} = -\frac{1}{2} \sum_{l_1 m_1} S^{-1}_{l_1m_1lm} a^{*}_{l_1m_1} + 
 \frac{1}{2} \sum_{l_1 m_1}N^{-1}_{l_1m_1lm} (d^{*}_{l_1m_1} - a^{*}_{l_1m_1}).
\end{equation}
To compute $\dot{p^r}_{lm}$ and $\dot{p^i}_{lm}$, the momentum derivatives corresponding to the real and imaginary part of $a_{lm}$ respectively, we make use of the following relations
\begin{equation}
 \frac{\pd \calH}{\pd a^r_{lm}} = 2 \Re \Big[ \frac{\pd \calH}{\pd a_{lm}} \Big] 
 \quad \text{and} \quad
 \frac{\pd \calH}{\pd a^i_{lm}} = -2 \Im \Big[ \frac{\pd \calH}{\pd a_{lm}} \Big],
\end{equation}
where $\Re$ and $\Im$ denote the real and imaginary part of the complex number, respectively.
Hamilton's equations for $(m_{10}, m^r_{11}, m^i_{11})$ are
\begin{equation}
 \dot{m}_{10} = \frac{\bar{p}_{10}}{\mu_{10}}, ~\dot \bar{p}_{10} = 
		-\frac{\partial \calH}{\partial m_{10}},
 ~\dot{m^r}_{11} = \frac{\bar{p^r}_{11}}{\mu^r_{11}}, ~\dot \bar{p^r}_{11} = 
                -\frac{\partial \calH}{\partial m^r_{11}},
 ~\dot{m^i}_{11} = \frac{\bar{p^i}_{11}}{\mu^i_{11}}, ~\dot \bar{p^i}_{11} = 
                -\frac{\partial \calH}{\partial m^i_{11}}.
\end{equation}
\begin{equation}
 \frac{\pd \calH}{\pd m^r_{11}} = 2\Re \Big[ \frac{\pd \calH}{\pd m_{11}} \Big]
 \quad \text{and} \quad
 \frac{\pd \calH}{\pd m^i_{11}} = -2\Im \Big[ \frac{\pd \calH}{\pd m_{11}} \Big],
\end{equation}
where $\frac{\pd \calH}{\pd m_{1N}}$ are given in terms of derivative of $\calH$ with respect to 
$A^{1N}_{ll+1}$ as
\begin{equation}\label{Pm1M_dot}
\frac{\pd \calH}{\pd m_{1N}} = 
2 \sum_{l} \frac{\pd A^{1N}_{l l+1}}{\pd m_{1N}} \frac{\pd \calH}{\pd A^{1N}_{ll+1}}.
\end{equation}
The factor of 2 in above expression appears because $A^{1N}_{l l+1} = A^{1N}_{l+1 l}$ for the dipole modulation model.
The time derivative of the momentum corresponding to $\alpha$ is obtained as
\begin{equation}\label{Palpha_dot}
 \dot{p}_{\alpha} = \frac{\pd \calH}{\pd \alpha} = \sum_l \Big [ \frac{\pd \calH}{\pd 
A^{10}_{ll+1}(l_p, \alpha) } \frac{\pd A^{10}_{ll+1}(l_p, \alpha)}{\pd \alpha} + 
\frac{\pd \calH}{\pd A^{11}_{ll+1}(l_p, \alpha) } \frac{\pd A^{11}_{ll+1}(l_p, \alpha)}{\pd \alpha}
\Big ].
\end{equation}
From \Eq{Pm1M_dot} and \Eq{Palpha_dot}, we note that to compute derivatives of momentum 
corresponding to $m_{10}, m^r_{11}, m^i_{11}$ and $\alpha$, we need to know $\frac{\pd \calH}{\pd A^{1M}_{ll+1}}$. 

The time derivative of conjugate momentum corresponding to the BipoSH coefficient $A^{LN}_{ll'}$ is
\begin{equation}\label{PALMll_dot}
 \dot{p}^{LN}_{ll'} = - \frac{\pd \calH}{\pd A^{LN}_{ll'}} = 
 - \frac{1}{2} \frac{\pd ln |\bS|}{\pd A^{LN}_{ll'}} 
 - \frac{1}{2} \frac{\pd \ba^{\dagger}\bS^{-1} \ba }{\pd A^{LN}_{ll'}}. 
\end{equation}

Following is a detailed derivation of the expression for $\dot{p}^{LN}_{ll'}$ starting from \Eq{PALMll_dot}.
To explicitly evaluate \Eq{palm_dot_2} and to simplify \Eq{PALMll_dot} we use a Taylor series expansion of $\bS^{-1}$ up to first order in $\bO$ as 
 \begin{equation}
  \bS^{-1} = \bD - \bD^{-1} \bO \bD^{-1}.
\end{equation}
We make use of two standard results from matrix calculus
\begin{equation}\label{der_ln_matrix}
 \frac{\pd \ln |\bS|}{\pd A^{LN}_{ll'} } = Tr\Big[ \bS^{-1}\frac{\pd \bS}{\pd A^{LN}_{ll'}}  \Big],
\end{equation}
and 
\begin{equation}\label{der_inv_matrix}
 \frac{\pd \ba^{\dagger}\bS^{-1} \ba }{\pd A^{LN}_{ll'}} = 
 -Tr\Big[ (\bS^{-1}\ba) (\ba^{\dagger}\bS^{-1}) \frac{\pd \bS}{\pd A^{LN}_{ll'}}  \Big].
\end{equation}
The derivative of element of the covariance matrix with respect to BipoSH coefficient is
\begin{equation}\label{der_Sl1m1l2m2}
 \frac{\pd S_{l_1 m_1 l_2 m_2}}{\pd A^{LN}_{ll'} } = (-1)^{m_2} C^{LN}_{l_1 m_1 l_2 -m_2} 
\delta_{l_1l} \delta_{l_2l'}.
 \end{equation}
Using the Taylor series expansion, the element of the inverse of the covariance matrix is
\begin{equation}\label{Sl1m1l2m2}
  S^{-1}_{l' m' l m} = D_{l' m' l m} - D^{-1}_{l' m' l' m'}O_{l' m' l m} D^{-1}_{l m l m}.
\end{equation}
Using the aboe Taylor series expression and the derivative in \Eq{der_Sl1m1l2m2},
\Eq{der_ln_matrix} can be expressed as
\begin{equation}\label{der_ln_matrix_2}
 \frac{\pd \ln |\bS|}{\pd A^{LN}_{ll'}} = \sum_{m'} \frac{(-1)^{m'}}{D_{l'm'l'm'}} C^{LN}_{lm'l'-m'} 
\delta_{ll'} - \sum_{m', m} \frac{(-1)^{m'} O_{l'm'lm} }{D_{l'm'l'm'} D_{lmlm}} C^{LN}_{lml'-m'}.
\end{equation}
Using the derivative in \Eq{der_Sl1m1l2m2}, \Eq{der_inv_matrix} becomes
\begin{equation}\label{der_inv_matrix_2}
 \frac{\pd \ba^{\dagger}\bS^{-1} \ba }{\pd A^{LN}_{ll'}} = 
 - \sum_{m', m} (\bS^{-1}\ba)_{l'm'} (\ba^{\dagger}\bS^{-1})_{lm} (-1)^{m_1} C^{LN}_{lml'-m'}.
\end{equation}
The term in the above equation with the Taylor series approximation gives
\begin{eqnarray}\label{der_inv_matrix_complicated_term}
(\bS^{-1}\ba)_{l_1m_1} (\ba^{\dagger}\bS^{-1})_{l_2m_2} 
&=& \frac{a_{l_1 m_1} a^*_{l_2 m_2}}{D_{ l_1 m_1 l_1 m_1} D_{l_2 m_2 l_2 m_2} } 
- \frac{a_{l_1 m_1}}{D_{ l_1 m_1 l_1 m_1} D_{l_2 m_2 l_2 m_2} } \sum_{l'_2 m'_2} 
\frac{O_{l'_2 m'_2 l_2 m_2} a^*_{l'_2 m'_2}}{D_{l'_2 m'_2 l'_2 m'_2} } \nonumber\\
&-& \frac{a_{l_2 m_2}}{D_{ l_1 m_1 l_1 m_1} D_{l_2 m_2 l_2 m_2} } \sum_{l'_1 m'_1} 
\frac{O_{l'_1 m'_1 l_1 m_1} a^*_{l'_1 m'_1}}{D_{l'_1 m'_1 l'_1 m'_1} } \nonumber\\
&+& \sum_{l'_1 m'_1} \sum_{ l'_2 m'_2} \frac{O_{l_1 m_1 l'_1 m'_1} O_{l'_2 m'_2 l_2 m_2} } 
{D_{l'_1 m'_1 l'_1 m'_1} D_{l'_2 m'_2 l'_2 m'_2} } \frac{a_{l'_1 m'_1} a^*_{l'_2 m'_2}}
{D_{ l_1 m_1 l_1 m_1} D_{l_2 m_2 l_2 m_2} }.
\end{eqnarray}
In our computation, we replace the second and third term by its ensemble average and keep terms only 
upto first 
order in $\bO$.
\begin{equation}\label{der_inv_matrix_complicated_term_simplified}
(\bS^{-1}\ba)_{l_1m_1} (\ba^{\dagger}\bS^{-1})_{l_2m_2} 
= \frac{a_{l_1 m_1} a^*_{l_2 m_2}}{D_{ l_1 m_1 l_1 m_1} D_{l_2 m_2 l_2 m_2} } 
- \frac{O_{l_1 m_1 l_2 m_2}}{D_{ l_1 m_1 l_1 m_1} D_{l_2 m_2 l_2 m_2} }
- \frac{O_{l_2 m_2 l_1 m_1}}{D_{ l_1 m_1 l_1 m_1} D_{l_2 m_2 l_2 m_2}}.
\end{equation}
Eq. \eqref{der_inv_matrix_2} along with above simplification and \Eq{der_ln_matrix_2} leads to following 
expression 
for $\dot{p}^{LN}_{ll'}$
\begin{eqnarray}\label{PALMll_dot_2}
 \dot{p}^{LN}_{ll'} = &-& \frac{1}{2} \sum_{m'} \frac{(-1)^{m'}}{D_{l'm'l'm'}} C^{LN}_{lm'l'-m'} 
\delta_{ll'} \nonumber\\
 &-& \frac{1}{2} \sum_{m', m} \frac{(-1)^{m'} O_{l'm'lm} }{D_{l'm'l'm'} D_{lmlm}} C^{LN}_{lml'-m'}
 + \frac{1}{2} \sum_{m', m} \frac{(-1)^{m'} a_{l'm'} a^*_{lm} }{D_{l'm'l'm'} D_{lmlm}} 
C^{LN}_{lml'-m'}.
\end{eqnarray}
The time derivative of the conjugate momentum of $C_l$ is obtained from \Eq{PALMll_dot_2} by 
substituting $L = 0$ and $M = 0$ in the above equation (as $A^{00}_{ll}$ are related to the $C_l$)
\begin{equation}\label{eq_PA00ll_dot}
 \dot{p}^{00}_{ll'} = 
 \frac{2l + 1}{2A^{00}_{ll}} (\frac{\hat{A}^{00}_{ll'}}{A^{00}_{ll}} - 1),
 \quad \text{where} \quad 
 \hat{A}^{00}_{ll'}  = \sum_{m m'} a_{lm} a^{*}_{l'm'} C^{00}_{lml'm'}.
\end{equation}
By using relation between $A^{00}_{ll}$ and $C_l$, we get
\begin{equation}\label{eq_PCl_dot}
 \dot{p}_{C_l} = \frac{2l + 1}{2 C_l}(\frac{\hat{C}_l}{C_l} - 1), \quad \text{where} \quad 
 \hat{C}_l = \frac{1}{2l + 1} \sum_{m} |a_{lm}|^2.
\end{equation}
Equation \ref{eq_PCl_dot} is same as equation (25) in \cite{Taylor_Cl_with_HMC_2008}. Note that, in 
\Eq{eq_PA00ll_dot} and \Eq{eq_PCl_dot}, $A^{00}_{ll'}$ and $C_l$ are the variables of dynamics, 
whereas $\hat{A}^{00}_{ll'}$ and $\hat{C}_l$ are quantities obtained using other variables of 
dynamics, namely \{$a_{lm}$\}.

\section{Demonstration of method on simulated map using CoNIGS} \label{Demo_on_sim_map}
\begin{figure}[H]
\includegraphics[width=0.8\linewidth, 
center]{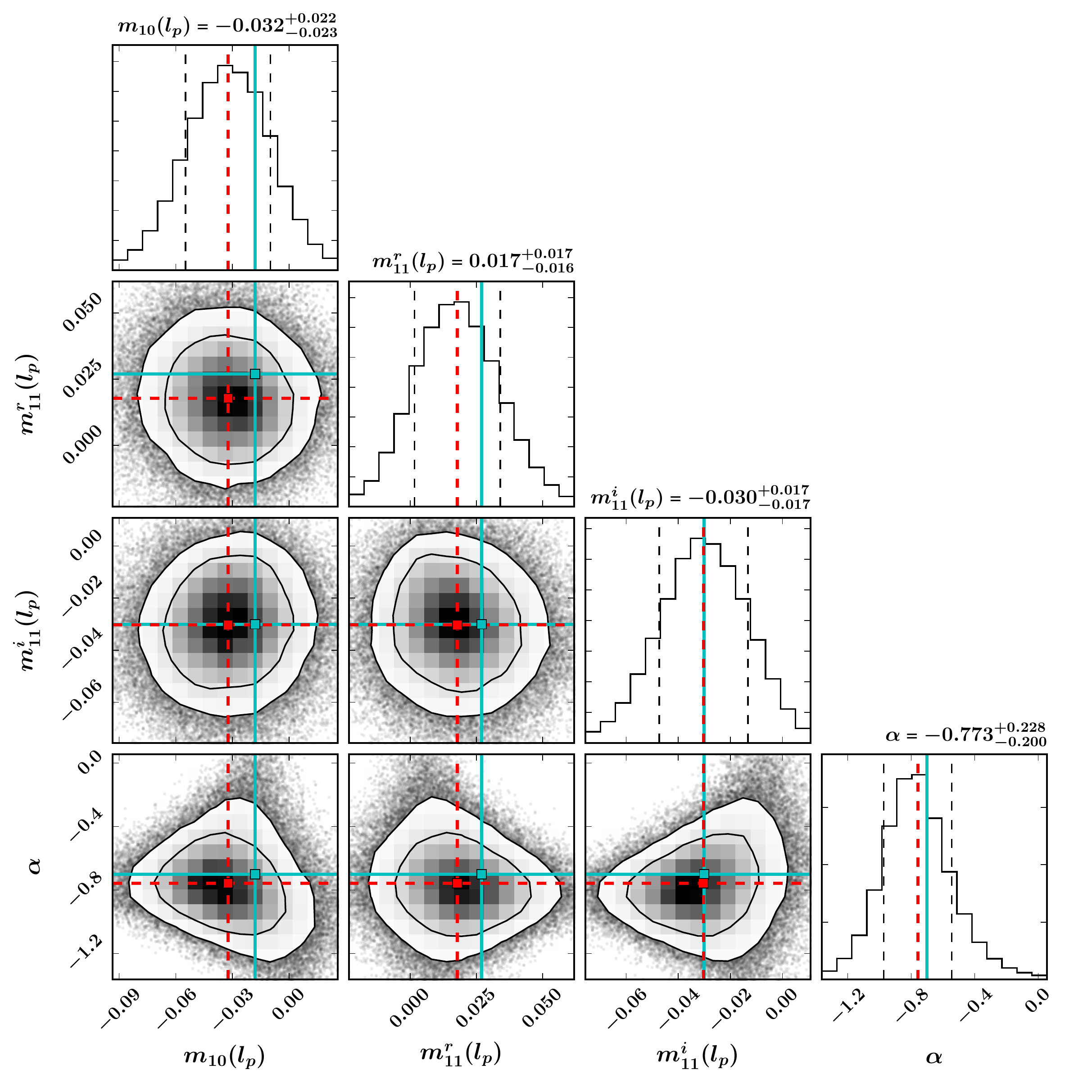}
\caption{The plot shows the joint and marginalized distributions of parameters $m_{10}(l_p = 32)$, $m^{r}_{11}(l_p = 32)$, $m^{i}_{11}(l_p = 32)$ and $\alpha$ for simulated map. Cyan line marks the input 
value of respective parameter. Red dashed line indicates the mean of the sampled distribution. In 2D distributions, contours show 
regions of distribution containing 68\% and 90\% samples.
In 1D distributions, black dashed lines mark 16 and 84 percentiles of the distribution. The title above 
each histogram shows the median value and the 16 and 84 percentiles for the parameter. Respective 
mean and standard deviation of the parameters ($m_{10}(l_p), m^{r}_{11}(l_p), m^{i}_{11}(l_p), 
\alpha$) are $(-0.032, 0.023)$, $(0.018, 0.016)$, $(-0.030, 0.017)$, $(-0.75, 0.24)$.}
\label{fig_conigs_param_triangle}
\end{figure}

\begin{figure}[H]
\centering
\subfigure[ ]{\label{fig_conigs_A}  
\includegraphics[width=0.45\textwidth]{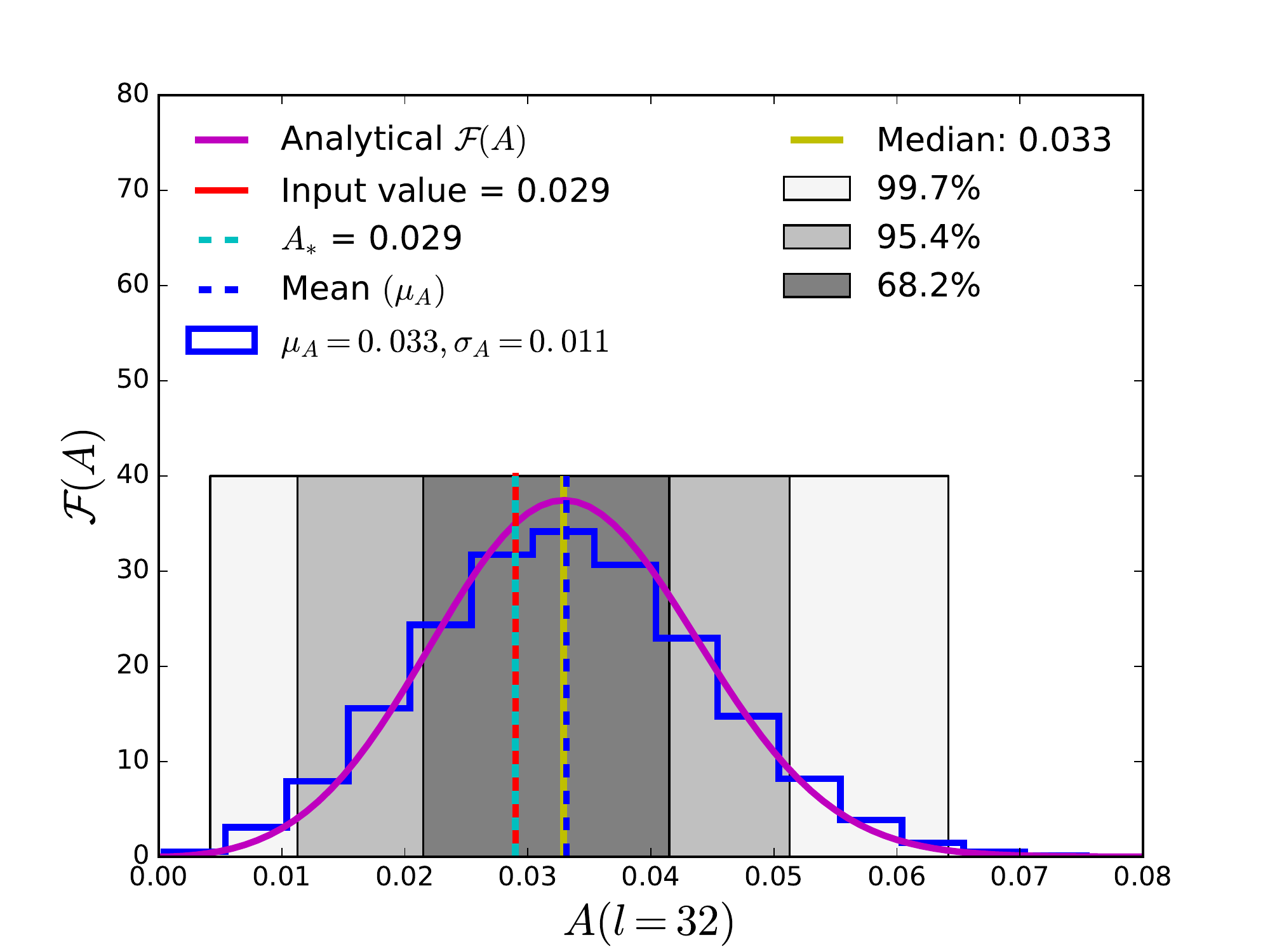}}
\subfigure[ ]{\label{fig_conigs_theta_phi} 
\includegraphics[width=0.45\textwidth]{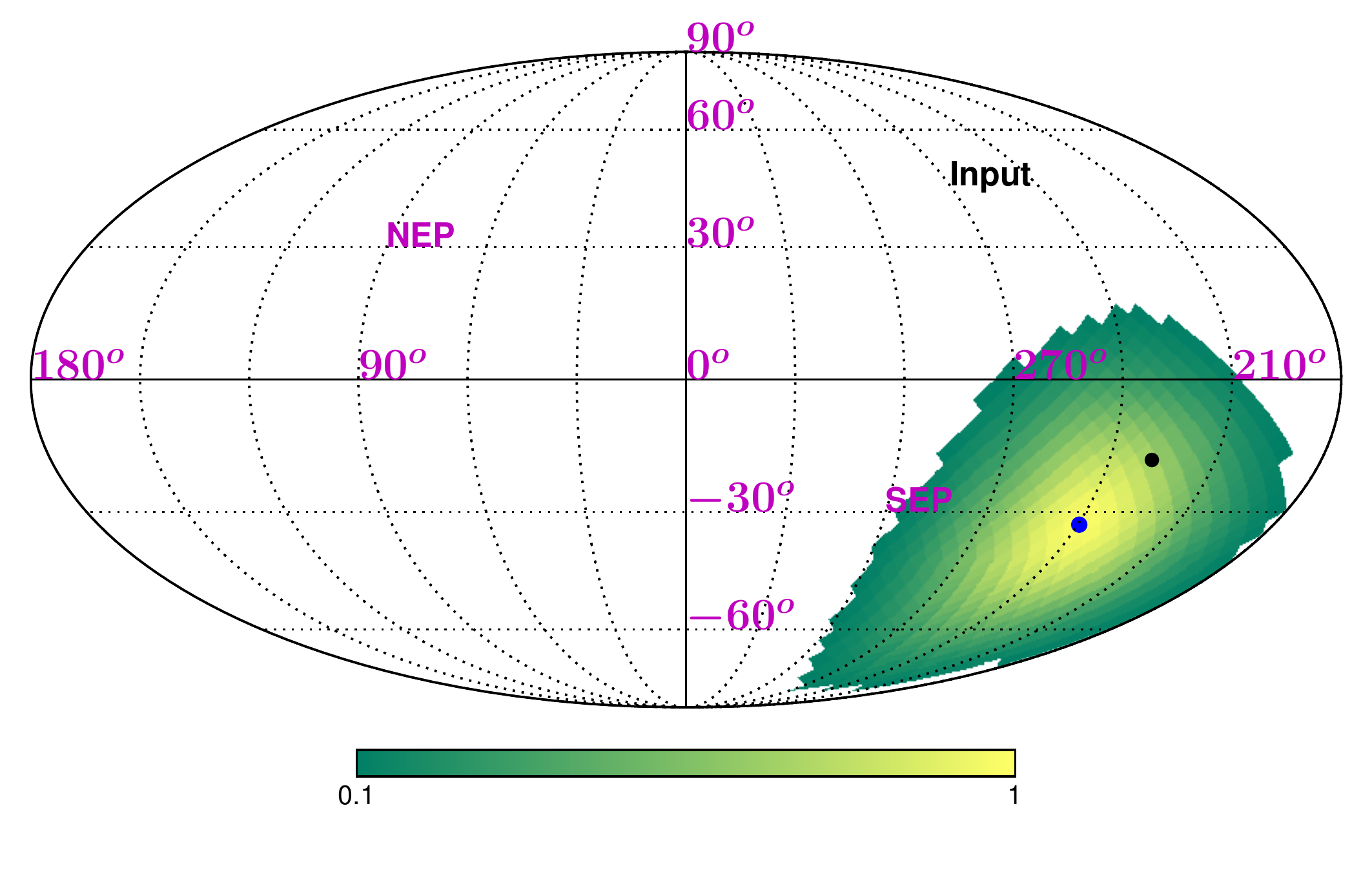}}
\caption{Figure in left panel shows the distribution the dipole amplitude at pivot multipole $l = 32$. The red line marks the input value of $A(l_p = 32)$ for the map, and blue dashed line indicates the mean of the distribution. $\mu_{A}$ and $\sigma_{A}$ are the mean and the standard deviation of $A(l_p = 32)$. The magenta curve shows the expected analytical distribution of $A(l_p = 32)$, based solely on the mean and standard deviations of $m_{10}(l_p)$, $m^{r}_{11}(l_p)$, $m^{i}_{11}(l_p)$ given in figure \ref{fig_conigs_param_triangle} (see \Eq{Prob_A_DM}). Our estimate of the dipole amplitude is given by $A_*(l_p = 32)$, the dipole amplitude corresponding to the mean values of $m_{10}(l_p), m^r_{11}(l_p), m^i_{11}(l_p)$, marked by dashed cyan line. Figure in right panel shows the distribution of $\theta_p$ and $\phi_p$ for simulated map in the galactic coordinate  system. $(\theta_p, \phi_p)$ values are binned using \texttt{HEALPix} \texttt{NSIDE} = 16 grid. Histogram so obtained is normalized with respect to peak and further smoothed by a Gaussian with standard deviation 3.7 degrees for presentation purpose. Black dot represents the direction of the input dipole in the simulation, $(l, b) = (228^o, -18^o)$. Blue dot represents maximum a posteriori estimate of the dipole direction $(\theta_p, \phi_p)$ having galactic coordinates $(l, b) = (239.5^o, -33.0^o)$.} \label{fig_conigs_A_theta_phi}
\end{figure}

\begin{figure}[H]
 \includegraphics[width=1.0\linewidth,center]
 {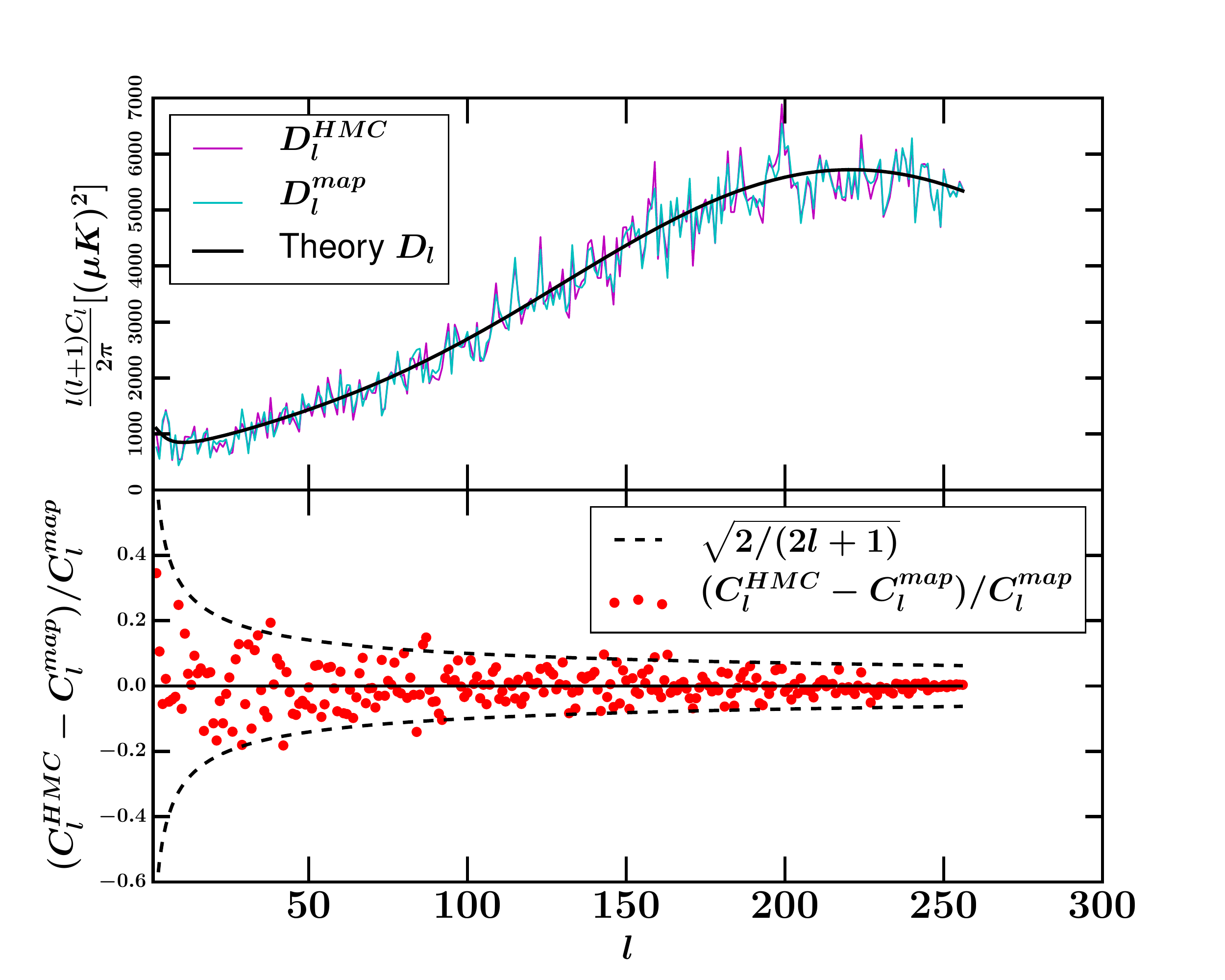}
\caption{The upper panel of the figure shows the maximum a posteriori estimate of the angular power 
spectrum from our analysis ($C^{HMC}_l$) and the angular power spectrum of the map realization 
($C^{map}_{l}$), for a simulated map. Black curve shows the \textit{theory} angular power 
spectrum used to generate the map. 
In bottom panel, we plot the relative difference between  $C^{HMC}_l$ and $C^{map}_{l}$ and compare it with the quantity $\sqrt{2/(2l + 1)}$, shown by black dashed curve. 
$D_l$ stands for $l(l+1)C_l/(2\pi)$.}
\label{fig_conigs_Cl}
\end{figure}

We test our algorithm on the simulated maps with the power law dipole modulation signal, in presence of anisotropic noise and masking. The power law dipole 
modulated CMB map is generated using \texttt{CoNIGS} algorithm \cite{Mukherjee_conigs}. We add the anisotropic noise \cite{Planck_FFP_2015} to this simulated map and do the analysis in presence of SMICA mask. 
We take $\{m_{10}(l_p = 32), m^r_{11} (l_p = 32), m^i_{11}(l_p = 32)\} = \{-0.018, 0.027, -0.030\}$ and $\alpha = -0.7$. The values of $\{m_{10}(l_p), m^r_{11}(l_p), m^i_{11}(l_p)\}$ are chosen so as to mimic the direction of the observed CHA based on the BipoSH-MVE estimate for the SMICA map \cite{Planck_2015_isotropy}. 
We do the analysis over the multipole range $l = 2 - 256$. In this way we confirm that we recover the signal consistent with that injected in the simulated map. Results of such an HMC run on one map are presented in figure \ref{fig_conigs_param_triangle}, figure \ref{fig_conigs_A}, figure \ref{fig_conigs_theta_phi} and figure \ref{fig_conigs_Cl}.

Histograms in figure \ref{fig_conigs_param_triangle} represent the distributions of harmonic 
parameters of dipole, $m_{10}(l_p), m^r_{11}(l_p)$ and $m^i_{11}(l_p)$ at the pivot multipole $l_p = 32$, and the power law index $\alpha$. These distributions are obtained using Monte Carlo chain of $10^5$ samples after discarding $5 \times 10^4$ samples as Burn-In. The set of 1D histograms, marginalized 
distributions of the parameters, in figure \ref{fig_conigs_param_triangle} show that the sampled distribution is consistent with the input values of the respective parameters for the map. We test the algorithm on different map realizations. We find that the mean values of $m_{10}, m^r_{11}$ and $m^i_{11}$ samples vary fairly randomly around the values assumed in simulation. This suggests that our 
estimates do not have any bias.

From \Eq{var_m10} for the variance of $m_{10}$, we know that the variance depends on the range of multipoles over which the dipole modulation signal exists and the power law index. This information is not available prior to the analysis of real data. For the simulated map, $\alpha = -0.7$ and $l = 2$ to $l = 256$. This leads to the following: $\sigma_{m_{10}} = 0.023$ and $\sigma_{m^r_{11}} = \sigma_{m^i_{11}} = 0.016$. The corresponding figures given in figure 
\ref{fig_conigs_param_triangle} are in agreement with these estimates. The standard deviations of 
sampled distributions are close to the values expected from the analytical arguments. Using 
$\{m_{10}(l_p = 32), m^r_{11} (l_p = 32), m^i_{11}(l_p = 32)\}$ we obtain the samples of $A(l_p = 32), \theta_p, \phi_p$ using \Eq{m1M_to_Athephi}. Figure \ref{fig_conigs_A} shows the distribution of $A(l_p = 32)$ obtained from the harmonic space variables shown in figure \ref{fig_conigs_param_triangle}. Figure \ref{fig_conigs_A} indicates that our method does recover the input value present in the simulated map. Figure \ref{fig_conigs_theta_phi} shows the distribution of $\theta_p, \phi_p$ samples. We see that our best-fit direction is close to the dipole modulation direction introduced in the simulated map.

A significant fraction of our computation is expended in incorporating the estimation of the map angular 
power spectrum. A summary statistics of $C_l$ distribution is given in figure \ref{fig_conigs_Cl}. 
Using the Monte Carlo chain of $C_l$ samples, we obtain the maximum a posteriori estimate of 
$C_l$ $(C^{HMC}_l)$ from our samples. We fit the analytical probability distribution of $C_l$ given 
in \cite{Percival:2006ss} to the histogram of $C_l$ samples. We use \texttt{SciPy} 
routine \texttt{scipy.optimise.curve\_fit} to do the fitting \cite{scipy_tools}. In figure 
\ref{fig_conigs_Cl} we compare $C^{HMC}_l$ with the angular power spectrum of the map realization, 
$(C^{map}_l)$. It is clear that we faithfully recover the estimate of angular power spectrum using 
our algorithm. 

\section{Establishing robustness to the choice of pivot multipole}
\label{alt_method}
\begin{figure}[H]
\includegraphics[width=0.8\linewidth, 
center]{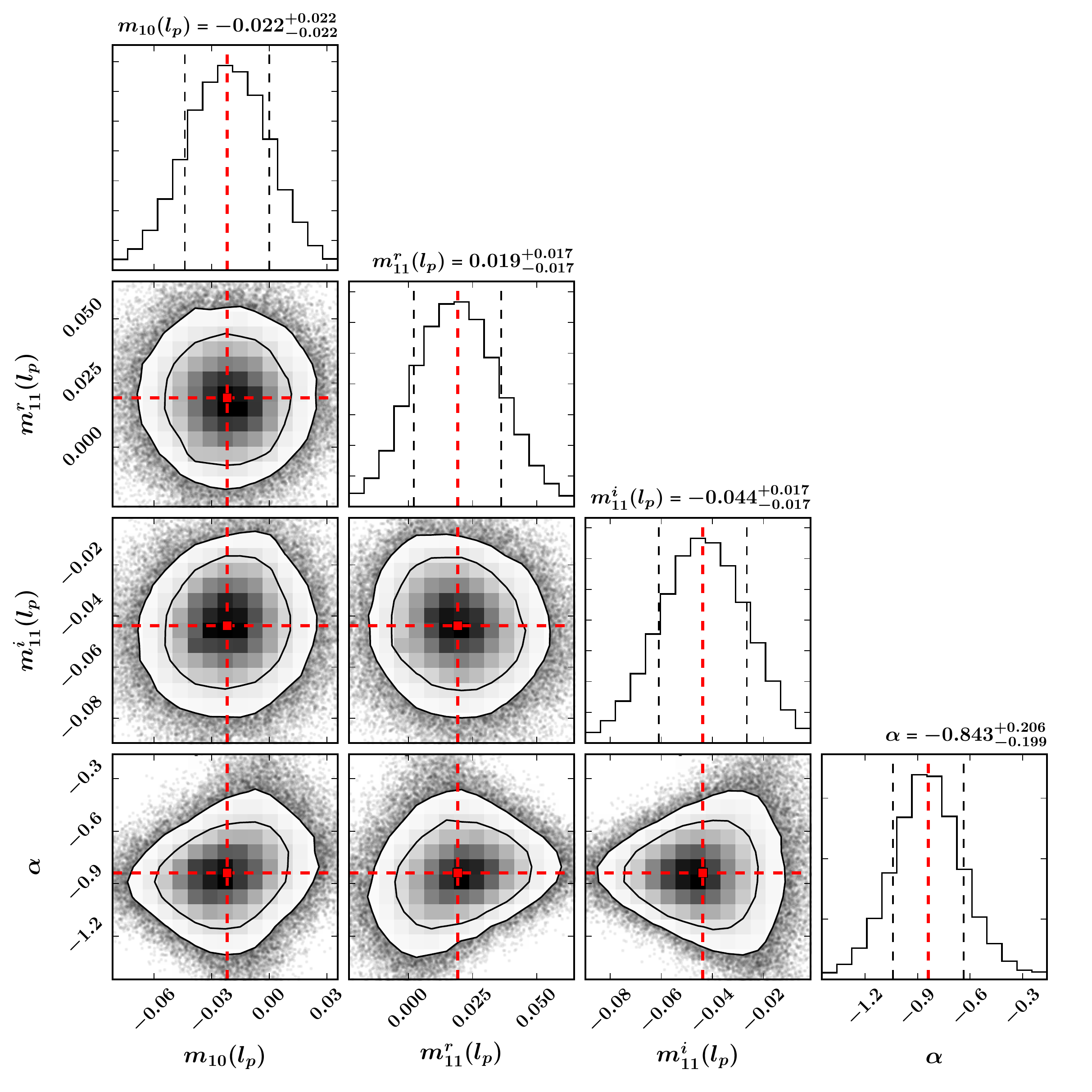}
\caption{The plot shows the joint and marginalized distributions of parameters $m_{10}(l_p = 32)$, 
$m^{r}_{11}(l_p = 32)$, $m^{i}_{11}(l_p = 32)$ and $\alpha$ for the SMICA map. Red dashed line indicates the mean of the sampled distribution. In 2D distributions, contours show 
regions of distribution containing 68\% and 90\% samples.
In 1D distributions, black dashed lines mark 16 and 84 percentiles of the distribution. The title above each histogram shows the median value and the 16 and 84 percentiles for the parameter. Respective mean and standard deviation of the parameters ($m_{10}(l_p), m^{r}_{11}(l_p), m^{i}_{11}(l_p), \alpha$) are $(-0.022, 0.022)$, $(0.019, 0.017)$, $(-0.044, 0.017)$, $(-0.84, 0.22)$.} 
\label{fig_smica_lp32_param_triangle}
\end{figure}

For the power law model analysis of SMICA map presented in the main body of the article, we chose $l = 16$ as the pivot multipole. To corroborate the answer we get for $l_p = 16$, in this appendix we provide the result for a different choice of pivot multipole, $l_p = 32$. The distributions of parameters $m_{10}(l_p = 32), m^r_{11}(l_p = 32), m^i_{11}(l_p = 32)$ and $\alpha$ are given in the figure \ref{fig_smica_lp32_param_triangle}. Taking $\alpha = -0.92$ and extrapolating from the results given in figure \ref{fig_smica_param_triangle} for $l_p = 16$, we expect $m_{10}(l = 32) = -0.023, m^r_{11}(l = 32) = 0.017, m^i_{11}(l = 32) = -0.043$. These values are consistent with the estimates of corresponding parameters given in figure \ref{fig_smica_lp32_param_triangle}.  

\section{HMC analysis of CHA with fixed fiducial \texorpdfstring{$\Lambda$}{Lg}CDM \texorpdfstring{$C_l$}{Lg}}\label{Appendix_step_model_SMICA}
\begin{figure}[H]
\subfigure[ ]{\label{fig_smica_A_fixCl}  
\includegraphics[width=0.5\linewidth]{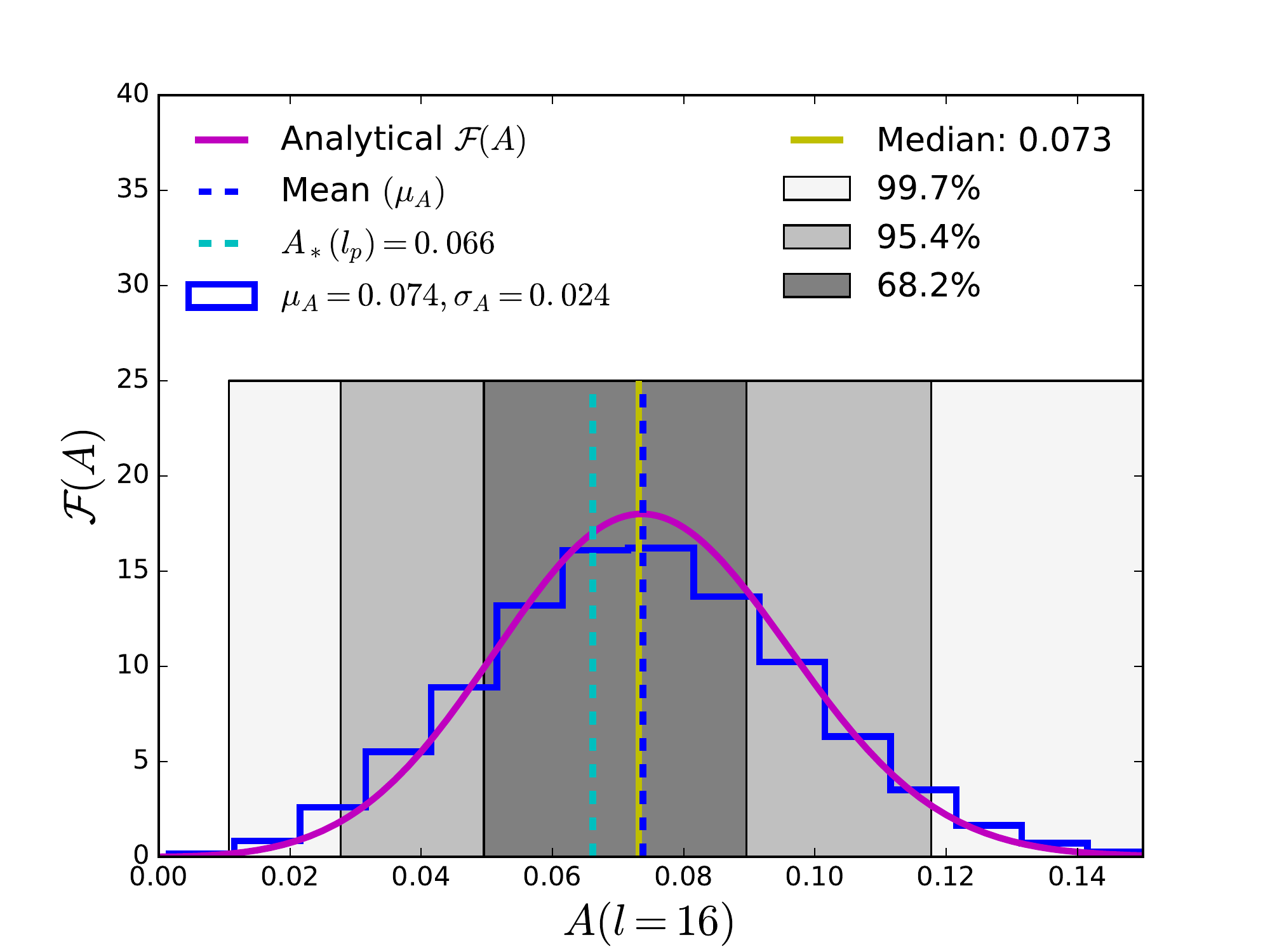}}
\subfigure[ ]{\label{fig_smica_thephi_fixCl} 
\includegraphics[width=0.5\linewidth]{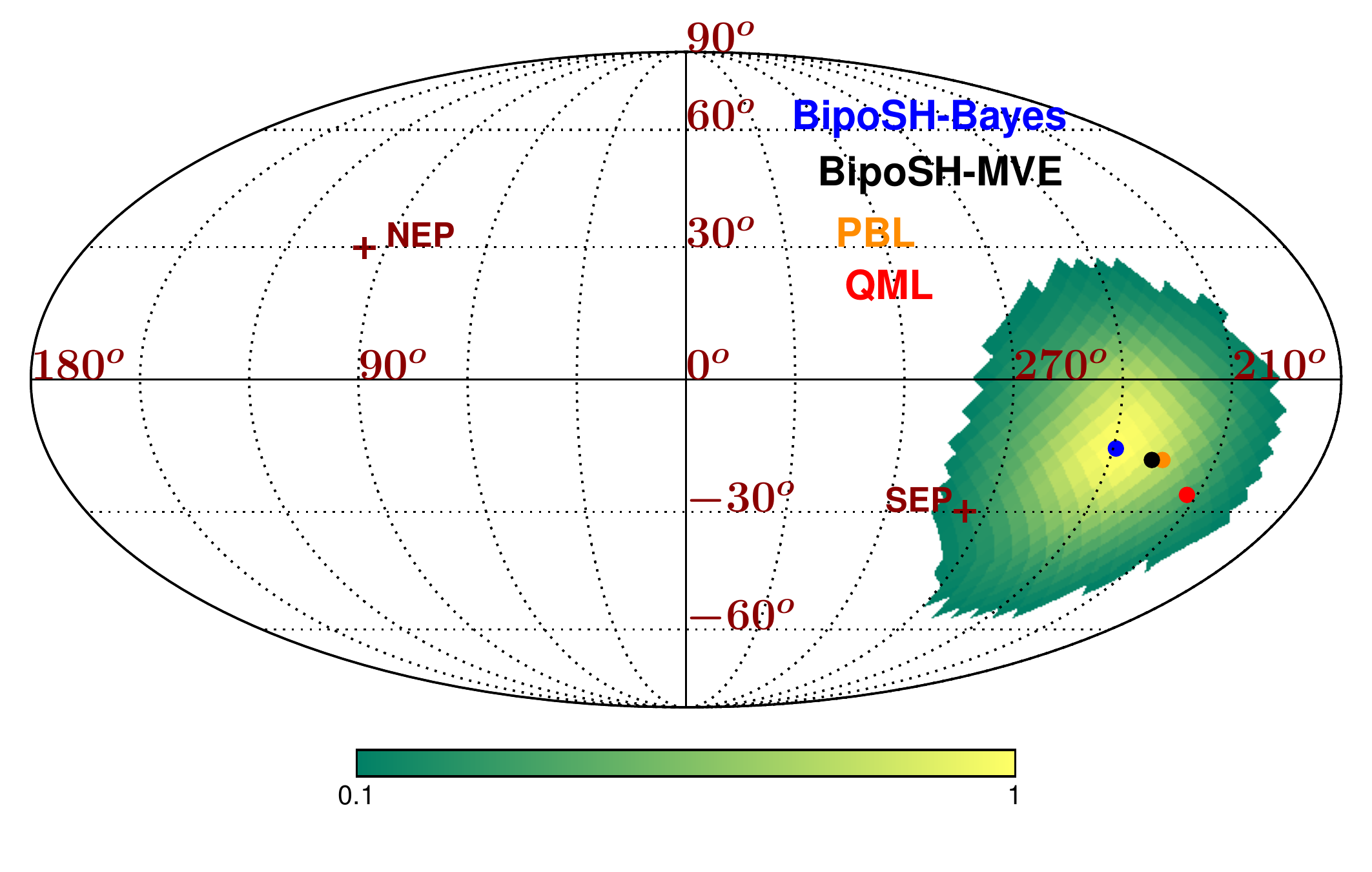}}
\caption{In left panel, the histogram is normalized probability distribution of $A(l_p)$, for fix $C_l$. Shaded regions mark the  68\%, 95\% and 99.7\% areas under the histogram. Dashed blue vertical line marks the mean  $(\mu_{A})$ of the distribution, $\sigma_{A}$ is the standard deviation of the distribution. Cyan dashed line marks $A_*$, the dipole amplitude corresponding to the mean values of $m_{10}, m^r_{11}, m^i_{11}$. The magenta curve shows the analytical distribution of $A$ given in \Eq{Prob_A_DM} with $A_* = 0.066$. In right panel, the distribution of $\theta_p$ and $\phi_p$ for SMICA map for fix $C_l$. $(\theta_p, \phi_p)$ values are binned using \texttt{HEALPix} \texttt{NSIDE} = 16 grid, which is further smoothed by a gaussian with standard deviation 2.8 degree for representation purpose. We adopt the galactic coordinate system for this plot. Blue dot represents the direction specified by the peak of the 2D histogram: $(l, b) = (239.2^o, -15.4^o)$. Also shown are the estimates from  \cite{Planck_2015_isotropy}: (1) PBL estimate (orange) $(l, b) = (225^o, -18^o)$, (2) QML estimate (red) $(l, b) = (213^o, -26^o)$, and (3) BipoSH-MVE estimate (black) $(l, b) = (228^o, -18^o)$.} \label{fig_smica_Athephi_fixCl}
\end{figure}

From \Eq{DM_biposh} and \Eq{shape_factor} we see that the nature of $C_l$ affects the magnitude of $m_{1N}$. In the main body of this article we performed the joint analysis of $C_l$ and $m_{1N}$ and find that we recover the $C_l$ consistent with the one reported in the literature \cite{Planck_2015_Cl}. We also estimate the correction to $C_l$ due to second order term in the dipole modulation covariance matrix, which turns out to relatively small except at low multipole $l < 5$, below which correction is around 1\%. In this section, we provide the result of the sampling of $m_{10}(l_p), m^r_{11}(l_p), m^i_{11}(l_p), \alpha$ where the angular power spectrum $C_l$ is held fixed at the best-fit $\Lambda$CDM power spectrum provided by \textit{Planck} \cite{Planck_2015_Cl}. The distribution of $A(l_p)$ and that of $(\theta, \phi)$ given in figure \ref{fig_smica_A_fixCl} and figure \ref{fig_smica_thephi_fixCl} respectively are similar to those given in figure \ref{fig_smica_A} and figure \ref{fig_smica_theta_phi} with $C_l$ as variable.

\section{Establishing robustness of the evidence ratio against prior choice}\label{Prior_choices}
In section \ref{model_comp}, we discussed a prior, say, \textit{Prior-1}, which is spherically symmetric and has a constant density within the sphere of radius $R(\alpha)$. The angle marginalized prior density of the amplitude for Prior-1 goes as $r^2$. Here, we explore two other priors, \textit{Prior-2} for which the angle marginalized amplitude prior density goes as $r$ and \textit{Prior-3} for which angle marginalized amplitude prior density is constant with respect to $r$.

Prior-2 is spherically symmetric, centered at $(0,0,0)$ and is inversely proportional to the separation from $(0,0,0)$ till $R(\alpha)$ and zero afterward. The normalized form of such a prior is
\begin{equation}\label{prior2}
 \Pi_{2}(w_x, w_y, w_z|\alpha, DM) = 
 \begin{cases}
    \frac{1}{2 \pi R^2(\alpha) \sqrt{w^2_x + w^2_y + w^2_z}} & \text{if } \sqrt{w^2_x + w^2_y + w^2_z} \leq 
R(\alpha) \\
    0              & \text{otherwise}.
\end{cases}
\end{equation}
With Prior-2, the angle marginalized prior density of the amplitude is
\begin{equation}\label{marg_prior2}
 \Pi_{2}(r|\alpha, DM)  = \frac{2 r}{R(\alpha)^2} \quad \text{for } r \leq R(\alpha).
\end{equation}
Prior-3 is spherically symmetric, centered at $(0,0,0)$ and is inversely proportional to the square of separation from $(0,0,0)$ till $R(\alpha)$. \textit{Prior-3} is again zero outside the sphere of radius $R(\alpha)$. The normalized form of such a prior is
\begin{equation}\label{prior3}
 \Pi_{3}(w_x, w_y, w_z|\alpha, DM) = 
 \begin{cases}
    \frac{1}{4 \pi R(\alpha) (w_x^2 + w_y^2 + w_z^2)} & \text{if } \sqrt{w_x^2 + w_y^2 + w_z^2} \leq 
R(\alpha) \\
    0              & \text{otherwise}.
\end{cases}
\end{equation}
The prior density $\Pi_3$ has the property that it gives uniform probability density to the amplitude of the modulation dipole because the angle marginalized prior density on the amplitude is
\begin{equation}\label{marg_prior3}
 \Pi_{3}(r|\alpha, DM) = \frac{1}{R(\alpha)} \quad \text{for } r \leq R(\alpha).
\end{equation}

The prior densities $\Pi_2$ and $\Pi_3$ diverge as $(w_x,w_y,w_z) \rightarrow (0,0,0)$. Hence we can not use the formula for the SDDR given in \Eq{SD_ratio_1} directly to evaluate the Bayes factor. In practice,  the numerator and denominator of the SDDR are computed by estimating the densities in a small but finite volume around the nested point.  In this section we present an attempt to evaluate the posterior to prior density ratio at a point $(x,y,z) = (\epsilon, \epsilon, \epsilon)$ close to $(x,y,z) = (0,0,0)$, where $\epsilon$ is small enough that it cannot be distinguished from zero at the level of the noise. This essentially amounts to cutting off the divergence in the prior at $r=\epsilon$. Modifying \Eq{SD_ratio_1}, the Bayes factor in favour of the SI model is
\begin{equation}\label{SD_ratio_2}
B_{SI-DM} = 
\frac{\calP_{C_l}(w_x = \epsilon, w_y = \epsilon, w_z = \epsilon,\alpha |d,DM)}
{\Pi(w_x = \epsilon, w_y = \epsilon, w_z = \epsilon, \alpha | DM)}.
\end{equation}
For Prior-1, the above expression is evaluated at $(w_x, w_y, w_z) = (0,0,0)$ and is exactly the SDDR in favor of SI model. We estimate the posterior density $\calP_{C_l}(w_x = \epsilon, w_y = \epsilon, w_z = \epsilon,\alpha |d,DM)$, which appears at the numerator of \Eq{SD_ratio_2}, using the Monte-Carlo samples and need a finite bin-width (around $3 \times 10^{-2}$) to get the estimate of posterior at any point. As a result of this binning, the posterior estimate does not change over the scale of the bin and is independent of $\epsilon$ as long as $(w_x, w_y, w_z) = (0,0,0)$ and $(w_x, w_y, w_z) = (\epsilon, \epsilon, \epsilon)$ are within the same bin. 
We obtain the ratio of posterior density and prior density in the grid containing the point $(x, y, z) = (\epsilon, \epsilon, \epsilon)$ with $\epsilon = 10^{-3}$. SDDR thus obtained for Prior-2 and Prior-3 is given in Table ~\ref{tab:title4}. 

From \Eq{SD_ratio_2}, SDDR depends on $\alpha$. However, at the nested point, which reduces the power law dipole modulation model to SI model, the likelihood does not depend on $\alpha$ and the SDDR also remains roughly constant over the range of $\alpha$. In Table ~\ref{tab:title4} we quote the value of SDDR averaged over $\alpha$.

\begin{table}[H]
\captionsetup{justification=centering}
\captionof{table}{Savage-Dickey density ratio (SDDR) for different prior densities}\label{tab:title4} 
\centering
\begin{tabular}{ |m{3cm}|m{3.0cm}|m{3.0cm}|}
\hline
 Prior choice & $(w_x, w_y, w_z)$ & SDDR \\
\hline
\hline
 $\Pi_2$ & $(\epsilon, \epsilon, \epsilon), \epsilon = 10^{-3}$   & 0.4 \\
\hline
 $\Pi_3$ & $(\epsilon, \epsilon, \epsilon), \epsilon = 10^{-3}$   & 0.5 \\
\hline
\end{tabular}
\end{table}


\label{Bibliography}

\bibliographystyle{JHEP}
\bibliography{hmc_reference}
\end{document}